\documentclass[aps,preprint,showpacs,showkeys]{revtex4}%
\usepackage{amsfonts}
\usepackage{amsmath}
\usepackage{amssymb}
\usepackage{graphicx}%
\providecommand{\U}[1]{\protect\rule{.1in}{.1in}}

\newtheorem{theorem}{Theorem}

\begin{document}
\preprint{ }
\title[ ]{The mathematical and geometrical structure of the spacetime and the concept of
unification, matter and energy}
\author{Diego Julio Cirilo-Lombardo\footnote{{\em e-mail}: diego777jcl@gmail.com}}
\author{}
\affiliation{International Institute of Physics, Federal University of Rio Grande do Norte,
59078-400 - Natal-RN, Brazil}
\affiliation{Bogoliubov Laboratory of Theoretical Physics, Joint Institute for Nuclear
Research, 141980, Dubna (Moscow region), Russian Federation}
\author{}
\affiliation{}

\begin{abstract}
Geometrical analysis of a new type of Unified Field Theoretical models follow
the guidelines of previous works\ of the authors is presented. These new
unified theoretical models are characterized by an underlying hypercomplex
structure, zero non-metricity and the geometrical action is determined
fundamentally by the curvature\textit{ }provenient of the breaking of symmetry
of a group manifold in higher dimensions. This mechanism of
Cartan-MacDowell-Mansouri type, permits us to construct geometrical actions of
determinantal type leading a non topological physical Lagrangian due the
splitting of a reductive geometry. Our goal is to take advantage of the
geometrical and topological properties of this theory in order to determine
the minimal group structure of the resultant spacetime Manifold able to
support a fermionic structure. From this fact, the relation between
antisymmetric torsion and Dirac structure\ of the spacetime is determined and
the existence of an important contribution of the torsion to the giromagnetic
factor of the fermions, shown. Also we resume and analyze previous
cosmological solutions in this new UFT where, as in our work [Class. Quantum
Grav. 22 (2005) 4987--5004] for the non abelian Born-Infeld model, the Hosoya
and Ogura ansatz is introduced for the important cases of tratorial, totally
antisymmetric and general torsion fields. In the case of spacetimes with
torsion the real meaning of the spin-frame alignment is find and the question
of the minimal coupling is discussed.
\end{abstract}
\maketitle
\tableofcontents

\section{Motivation and summary of the results}

From long time ago in the history of the modern theoretical physics the
possibility of the unification of all fundamental forces have been treated
from the mathematical and theoretical point of view. Several models,
formulations and sophisticated mathematical tools were used in order to solve
the intricate puzzle of to conciliate gravity with the other fundamental
forces of the nature: electromagnetic, weak and strong. Although many attempts
appear, this issue is still without concrete solution, as is the string theory
the typical case. In string theory is common the claiming about to be the
consistent solution of the unification trouble but, beside particular
formulations, \ the theoretical and conceptual environment joined with an
obscure mathematical basis put certainly in doubt the affirmative acceptation
of such claim.

As was pointed out by us in later works [1,2], the cornerstone of the problem
is where to start conceptually to reformulate the theoretical arena where the
fundamental unified theory will be placed, and where the geometry is the
unifying essence. According to Mach spacetime doesn't exists without matter.
Then, two basic ideas immediately arise to fulfill the observation given by
Mach: the concept of dualistic or non-dualistic theories. In the first one the
simplest and economical description can be formulated in terms of the
gravitational field without torsion plus the energy momentum tensor that,
however, is added "by hand" in order to cover the lack of knowledge of a
fundamental structure of the space time giving the matter plus energy
distribution. In the second one there are not prescriptions for the
interaction of gravity with the "matter" fields because they are arising from
the same fundamental geometrical structure.

In previous works of the authors we present a new model of a non-dualistic
Unified Theory. The goal that we introduce firstly in our preliminary model in
[1], absolutely consistent from the mathematical and geometrical point of
view, is that was based in a manifold equipped with an \textit{underlying
hypercomplex structure and zero non-metricity}, that lead the important fact
that the Torsion of the space-time structure turns to be totally
antisymmetric. As is well known in the particular case of totally
antisymmetric torsion tensor this type of affine geometrical frameworks have
the geodesics and the minimal length equations equivalent, and the most
important is that is the only case that the equivalence principle is fulfilled
as was shown in [9,10] and we demonstrate also here.

The other goal that we introduce as main ingredient in [1,2] and here, is that
the specific form of our action is determined by \textit{the curvature from
the breaking of symmetry of a group manifold in higher dimensions} via a
Cartan-MacDowell-Mansouri mechanism [1,2]. This mechanism permits to construct
geometrical actions of determinantal type that, due the splitting of a
reductive geometry (as is the case of the group manifold treated here) via the
breaking to the higher dimensional group (i.e.: as is the typical case
SO$\left(  1,4\right)  \rightarrow$ SO$\left(  1,3\right)  \oplus
\mathbb{M}_{1,3}))$leads a non topological physical Lagrangian.

Following the guidelines of our last works [1,2,3], in this paper we complete
the previous analysis considering the same fundamental model of UFT. The
organization of the paper with the corresponding results is as follows: in
Section II the geometrical framework is introduced and the theoretical basis
of the model, based in a geometrical action that takes physical meaning
through a breaking of symmetry, is described. In Section III the dynamic
equations are analyzed and the geometrical and physical meaning are elucidated.

In Section IV we resume and analyze previous cosmological solutions in the New
UFT: as in our work [3] for the non abelian Born-Infeld model, the Hosoya and
Ogura ansatz is introduced for the important cases of tratorial and totally
antisymmetric torsion. The real meaning of the spin-frame alignment in the
case with torsion is find. Also, we explicitly show that, contrarily to the
case of the Poincare theory of gravitation (see reference [4]), the
possibility in our Theory of the co-existence of both types of torsion in
cosmological spacetimes certainly exists.

Section V is the most important in the sense that the fermionic structure of
the spacetime is described and the possibility of geometrical unification
realized: a unified theory of QED\ and GR can be derived from P(G,M), the
Principal Fiber Bundle of frames over the 4D spacetime manifold with G as its
structure group. In the subsections, the action of the UFT\ is analyzed from
the group-theoretical\ point of view considering the G-symmetry of the model.
In Section VI\ the derivation of the Dirac equation from the G-manifold, the
relation between the electromagnetic field/fermionic structure of the
spacetime and the contribution of the torsion to the gyromagnetic factor are
explicitly shown. However, the physical consequences are explained. Finally,
Section VII\ is devoted to discuss the cohomological interplay between the
fields involved in the spacetime structure and in VIII the concluding remarks
are given.

\section{The space-time manifold and the geometrical action}

The starting point is an hypercomplex construction of the (metric compatible)
space-time manifold [1].%

\begin{equation}
M,g_{\mu\nu}\equiv e_{\mu}\cdot e_{\nu} \tag{1}%
\end{equation}
where for each point $\in M$ $\exists$ a local space affine $A.$The connection
over $A,$ $\ \widetilde{\Gamma}$ , define a generalized affine connection
$\Gamma$ on $M$ specified by $\left(  \nabla,K\right)  $ where $K$ is an
invertible $\left(  1,1\right)  $ tensor over $M.$ We will demand that the
connection is compatible and rectilinear%
\begin{equation}
\nabla K=KT,\text{ }\nabla g=0 \tag{2}%
\end{equation}

where $T$ is the torsion, and $g$ (the space-time metric, used for to raise
and to low indices and determines the geodesics) is preserved under parallel
transport. This generalized compatibility condition ensures that the affine
generalized connection $\Gamma$ maps autoparallels of $\Gamma$ on $M$ in
straight lines over the affine space $A$ (locally)$.$ The first equation is
equal to the condition determining the connection in terms of the fundamental
field in the UFT\ non-symmetric. For instance, $K$ can be identified with the
fundamental tensor in the non-symmetric fundamental theory. This fact give us
the possibility to restrict the connection to an (anti) Hermitian theory.

The covariant derivative of a vector with respect to the generalized affine
connection is given by%
\begin{equation}
\left.  A^{\mu}\right.  _{;\nu}\equiv\left.  A^{\mu}\right.  ,_{\nu}%
+\Gamma_{\ \ \alpha\nu}^{\mu}A^{\alpha} \tag{3}%
\end{equation}%
\[
A_{\mu;\nu}\equiv\left.  A_{\mu}\right.  ,_{\nu}-\Gamma_{\ \ \mu\nu}^{\alpha
}A_{\alpha}%
\]
The generalized compatibility condition (2) determines the 64 components of
the connection by the 64 equations as follows%
\begin{equation}
K_{\mu\nu;\alpha}=K_{\mu\rho}T_{\ \ \nu\alpha}^{\rho}\text{ where }%
T_{\ \ \nu\alpha}^{\rho}\text{ }\equiv2\Gamma_{\ \ \left[  \alpha\nu\right]
}^{\rho} \tag{4}%
\end{equation}
Notice that contraction of indices $\nu$ and $\alpha$ above in the first
equation (4), an additional condition over this hypotetic fundamental
(nonsymmetric) tensor $K$ is obtained%
\[
K_{\mu\alpha;\alpha}=0
\]
that, geometrically speaking is
\[
d^{\ast}K=0
\]
this is a current free condition over the tensor $K$ that can be exemplified
nicely with the prototype of non-symmetric fundamental tensor: $K_{\mu\nu
}=g_{\mu\nu}+$\ $f_{\mu\nu}$%
\[
d^{\ast}K=d^{\ast}g+d^{\ast}f\Rightarrow d^{\ast}f=0\text{ \ \ \ (current free
e.o.m.)}%
\]
where, however, playing $g_{\mu\nu}$ the role of spacetime metric and
\ $f_{\mu\nu}$ the role of electromagnetic field .

The metric is univoquely determined by the metricity condition that puts 40
restrictions on the partial derivatives of the metric
\begin{equation}
g_{\mu\nu,\rho}=2\Gamma_{\left(  \mu\nu\right)  \rho} \tag{5}%
\end{equation}
The space-time curvature tensor, that is defined in the usual way, has two
possible contractions: the Ricci tensor $R_{\mu\lambda\nu}^{\lambda}=R_{\mu
\nu}$ and the second contraction $R_{\lambda\mu\nu}^{\lambda}=2\Gamma
_{\ \ \lambda\left[  \nu,\mu\right]  }^{\lambda}$ is identically zero due the
metricity condition (2). In order to find a symmetry of the torsion tensor if
we denote the inverse of $K$ by $\widehat{K}$, $\widehat{K}$ is uniquely
specified by $\widehat{K}^{\alpha\rho}$ $K_{\alpha\sigma}=$ $K^{\alpha\rho}$
$\widehat{K}_{\alpha\sigma}=\delta_{\sigma}^{\rho}$. As was pointed out in
[1], inserting explicitly the torsion tensor as the antisymmetric part of the
connection in (4) and multiplying by $\frac{\widehat{K}^{\alpha\nu}}{2},$ this
results after straighforward computations in%
\begin{equation}
\left(  Ln\sqrt{-K}\right)  ,_{\mu}-\Gamma_{\left(  \mu\nu\right)  }^{\nu}=0
\tag{6}%
\end{equation}
where $K=\det$ $\left(  K_{\mu\rho}\right)  $. Notice that from expression (6)
we arrive to the following condition between the determinants $\ K$ and $\ g$:
$\frac{K}{g}=$cons$\tan$t. Now we can write
\begin{equation}
\Gamma_{\ \alpha\nu,\beta}^{\nu}-\Gamma_{\ \beta\nu,\alpha}^{\nu}%
=\Gamma_{\ \nu\beta,\alpha}^{\nu}-\Gamma_{\ \nu\alpha,\beta}^{\nu} \tag{7}%
\end{equation}
due the fact that the first term of is the derivative of an scalar. Then, the
torsion tensor has the symmetry%
\begin{equation}
T_{\ \nu\left[  \beta,\alpha\right]  }^{\nu}=T_{\ \nu\left[  \alpha
,\beta\right]  }^{\nu}=0 \tag{8}%
\end{equation}

That means that the trace of the torsion tensor defined as $T_{\ \nu\alpha
}^{\nu}$, is the gradient of a scalar%
\[
T_{\alpha}=\nabla_{\alpha}\phi
\]
The second important point is the following: let us consider [1] the extended
curvature%
\begin{equation}
\mathcal{R}_{\mu\nu}^{ab}=R_{\mu\nu}^{ab}+\Sigma_{\mu\nu}^{ab} \tag{9}%
\end{equation}
with%
\begin{equation}
R_{\mu\nu}^{ab}=\partial_{\mu}\omega_{\nu}^{ab}-\partial_{\nu}\omega_{\mu
}^{ab}+\omega_{\mu}^{ac}\omega_{\nu c}^{\ \ b}-\omega_{\nu}^{ac}\omega_{\mu
c}^{\ \ b} \tag{10}%
\end{equation}%
\[
\Sigma_{\mu\nu}^{ab}=-\left(  e_{\mu}^{a}e_{\nu}^{b}-e_{\nu}^{a}e_{\mu}%
^{b}\right)
\]
The second important point is the following: let us consider [1] the extended
curvature[8]%
\begin{equation}
\mathcal{R}_{\mu\nu}^{ab}=R_{\mu\nu}^{ab}+\Sigma_{\mu\nu}^{ab} \tag{9}%
\end{equation}
with [8]%
\begin{equation}
R_{\mu\nu}^{ab}=\partial_{\mu}\omega_{\nu}^{ab}-\partial_{\nu}\omega_{\mu
}^{ab}+\omega_{\mu}^{ac}\omega_{\nu c}^{\ \ b}-\omega_{\nu}^{ac}\omega_{\mu
c}^{\ \ b} \tag{10}%
\end{equation}%
\[
\Sigma_{\mu\nu}^{ab}=-\left(  e_{\mu}^{a}e_{\nu}^{b}-e_{\nu}^{a}e_{\mu}%
^{b}\right)
\]
We assume here $\omega_{\nu}^{ab}$ a $SO\left(  d-1,1\right)  $ connection and
$e_{\mu}^{a}$ is a vierbein field. The eqs.(9,10) can be obtained, for
example, using the formulation that was pioneering introduced in seminal works
by E. Cartan long time ago [1] . Is well known that in such an formalism the
gravitational field is represented as a connection one form associated with
some group which contains the Lorentz group as subgroup. The typical example
is provided by the $SO\left(  d,1\right)  $ de Sitter gauge theory of gravity.
In this specific case, the $SO\left(  d,1\right)  $ the gravitational gauge
field $\omega_{\mu}^{AB}=-\omega_{\mu}^{BA}$ is broken into the $SO\left(
d-1,1\right)  $ connection $\omega_{\mu}^{ab}$ and the $\omega_{\mu}%
^{da}=e_{\mu}^{a}$ vierbein field, with the dimension $d$ fixed. Then, the de
Sitter (anti-de Sitter) curvature
\begin{equation}
\mathcal{R}_{\mu\nu}^{AB}=\partial_{\mu}\omega_{\nu}^{AB}-\partial_{\nu}%
\omega_{\mu}^{AB}+\omega_{\mu}^{AC}\omega_{\nu C}^{\ \ B}-\omega_{\nu}%
^{AC}\omega_{\mu C}^{\ \ B} \tag{11}%
\end{equation}
splits in the curvature (9). At this point, our goal is to enlarge the group
structure of the spacetime Manifold of such manner that the curvature (11),
obviously after the breaking of symmetry, permits us to define the geometrical
Lagrangian of the theory as
\begin{equation}%
\begin{array}
[c]{ccc}
& L_{g}=\sqrt{det\mathcal{R}_{\ \mu}^{a}\mathcal{R}_{a\nu}} & =\sqrt
{detG_{\mu\nu}}%
\end{array}
\tag{12}%
\end{equation}

where we have been defined the following geometrical object%
\begin{equation}
\mathcal{R}_{\ \mu}^{a}=\lambda\left(  e_{\ \mu}^{a}+f_{\ \mu}^{a}\right)
+R_{\ \mu}^{a}\ \ \ \ \ \ \left(  M_{\ \mu}^{a}\ \equiv e_{\ }^{a\nu}M_{\nu
\mu}\right)  \ \tag{12}%
\end{equation}
where $f_{\ \mu}^{a}$ (in sharp contrast to $e_{\ \mu}^{a}$) carry the
following symmetry:
\[
e_{\ a\mu}f_{\ \nu}^{a}=f_{\mu\nu}=-f_{\nu\mu}.
\]

The action will contains, as usual, $\mathcal{R}=det\left(  \mathcal{R}%
_{\ \mu}^{a}\right)  $ as the geometrical object that defines the dynamics of
the theory. The particularly convenient definition of $\mathcal{R}_{\ \mu}%
^{a}$ makes easy to establish the equivalent expression in the spirit of the
Unified theories developed time ago by Eddington, Einstein and Born and Infeld
for example:%
\begin{equation}
\sqrt{det\mathcal{R}_{\ \mu}^{a}\mathcal{R}_{a\nu}}=\sqrt{det\left[
\lambda^{2}\left(  g_{\mu\nu}+f_{\ \mu}^{a}f_{a\nu}\right)  +2\lambda
R_{\left(  \mu\nu\right)  }+2\lambda f_{\ \mu}^{a}R_{\left[  a\nu\right]
}+R_{\ \mu}^{a}R_{a\nu}\right]  } \tag{13}%
\end{equation}
where $R_{\mu\nu}=R_{\left(  \mu\nu\right)  }+R_{\left[  \mu\nu\right]  }$.

The important point to consider in this simple Cartan inspired model is that,
although a cosmological constant $\lambda$ is required, the expansion of the
action in four dimensions lead automatically the Hilbert-Einstein part when
$f_{\ \mu}^{a}=0$. Explicitly ($R=g^{\alpha\beta}R_{\alpha\beta}$)%
\begin{align}
S  &  =\int d^{4}x\left(  e+f\right)  \left\{  \lambda^{4}+\lambda^{3}\left(
R+f_{\ \mu}^{a}R_{\ a}^{\mu}\right)  +\frac{\lambda^{2}}{2!}\left[
R^{2}-R^{\mu\nu}R_{\mu\nu}+\left(  f_{\ \mu}^{a}R_{\ a}^{\mu}\right)
^{2}-f^{\mu\nu}f^{\rho\sigma}R_{\mu\rho}R_{\nu\sigma}\right]  +\right.
\tag{14}\\
&
\begin{array}
[c]{c}%
+\frac{\lambda}{3!}\left[  R^{3}-3RR^{\mu\nu}R_{\mu\nu}+2R^{\mu\alpha
}R_{\alpha\beta}R_{\ \ \mu}^{\beta}+\left(  f_{\ \mu}^{a}R_{\ a}^{\mu}\right)
^{3}-3\left(  f_{\ \mu}^{a}R_{\ a}^{\mu}\right)  f^{\mu\nu}f^{\rho\sigma
}R_{\mu\rho}R_{\nu\sigma}+2f^{\mu\nu}R_{\mu}^{\ \alpha}R_{\alpha\beta}%
R_{\ \nu}^{\beta}\right] \\
\left.  +det\left(  R_{\mu\nu}\right)  \right\}
\end{array}
\nonumber
\end{align}
Notice that the tetrad property was used here. \ In the remaining part of the
work, this property will be used or not, wherever the case.

\section{The dynamical equations}

In this case, the variation with respect to the metric remains the same as in
previous works (see [1] eq.(9)):$e.g.$:%
\[
\delta_{g}\sqrt{G}=\frac{\sqrt{G}}{2}\left(  G^{-1}\right)  ^{\mu\nu}%
\delta_{g}G=0
\]
$.$ The variation respect to the connection gives immediately%

\begin{equation}
\frac{\delta\sqrt{G}}{\delta\Gamma_{\ \ \mu\nu}^{\omega}}=\left\{
-\nabla_{\sigma}\left[  \sqrt{G}\left(  G^{-1}\right)  ^{\alpha\nu}%
\mathcal{R}_{\ \ \alpha}^{\sigma}\right]  \delta_{\ \ \omega}^{\mu}%
+\nabla_{\omega}\left[  \sqrt{G}\left(  G^{-1}\right)  ^{\alpha\nu}%
\mathcal{R}_{\ \ \alpha}^{\mu}\right]  +\sqrt{G}\left(  G^{-1}\right)
^{\alpha\nu}\mathcal{R}_{\ \ \alpha}^{\sigma}\Gamma_{\ \ \left[  \sigma
\omega\right]  }^{\mu}\right\}  \tag{15}%
\end{equation}
where the general form of the Palatini's identity have been used and%
\[
G_{\mu\nu}\equiv\mathcal{R}_{\ \mu}^{a}\mathcal{R}_{a\nu}%
\]
with the $\mathcal{R}_{\ \mu}^{a}\mathcal{\ }$from eq.(12). Defining
$\Sigma^{\nu\sigma}\equiv\sqrt{G}\left(  G^{-1}\right)  ^{\alpha\nu
}\mathcal{R}_{\ \ \alpha}^{\sigma}$ the above equation can be written in a
more suggestive form but due the variation with respect to the metric it is
identically zero (due the lack of energy momentum tensor) and the only
information, till know, to our disposal is through the antisymmetric part of
the variation with respect to the metric$\left(  \text{see (12) of
ref}.[1]\right)  $%
\begin{gather}
R_{\mu\nu}=-\lambda\left(  g_{\mu\nu}+f_{\mu\nu}\right) \nonumber\\
\Rightarrow R_{\left[  \mu\nu\right]  }=\left(  \nabla_{\alpha}+2T_{\alpha
}\right)  \left(  T_{\mu\nu}^{\alpha}+T_{\nu}\delta_{\mu}^{\alpha}-T_{\mu
}\delta_{\nu}^{\alpha}\right)  =-2\lambda f_{\mu\nu} \tag{16}%
\end{gather}
with $T_{\alpha}$ the trace of the torsion tensor. Now we have to explore the
role played by $f_{\mu\nu}:$

i) if $f_{\mu\nu}$ plays the role of the electromagnetic field, then, we have
a one-form vector potential which $f_{\mu\nu}$ is derived. Notice the
important fact that such an existence not necessarily can follows "a priori"
from the definition of $f_{\rho\tau}.$ This fact lead to the usual
Euler-Lagrange equations, where the variation is made with respect to the
electromagnetic potential $a_{\tau}$
\begin{equation}
\frac{\delta\sqrt{G}}{\delta a_{\tau}}=\nabla_{\rho}\left(  \frac
{\partial\sqrt{G}}{\partial f_{\rho\tau}}\right)  \equiv\nabla_{\rho
}\mathbb{F}^{\rho\tau}=0 \tag{17}%
\end{equation}
Explicitly
\begin{equation}
\nabla_{\rho}\left[  \frac{\lambda^{2}N^{\mu\nu}\left(  \delta_{\mu}^{\sigma
}\ f_{\ \nu}^{\rho}+\delta_{\nu}^{\sigma}\ f_{\ \mu}^{\rho}\right)
}{2\mathbb{R}}\right]  =0 \tag{18}%
\end{equation}
where $N^{\mu\nu}$ is given by expression (32) of ref.[1]. The set of
equations to solve for this particular case is
\begin{gather}
R_{\left(  \mu\nu\right)  }=\overset{\circ}{R}_{\mu\nu}-T_{\mu\rho
}^{\ \ \ \alpha}T_{\alpha\nu}^{\ \ \ \rho}=-\lambda g_{\mu\nu}\tag{19a}\\
R_{\left[  \mu\nu\right]  }=\left(  \nabla_{\alpha}+2T_{\alpha}\right)
\left(  T_{\mu\nu}^{\alpha}+T_{\nu}\delta_{\mu}^{\alpha}-T_{\mu}\delta_{\nu
}^{\alpha}\right)  =-\lambda f_{\mu\nu}\tag{19b}\\
\nabla_{\rho}\left[  \frac{\lambda^{2}N^{\mu\nu}\left(  \delta_{\mu}^{\sigma
}\ f_{\ \nu}^{\rho}+\delta_{\nu}^{\sigma}\ f_{\ \mu}^{\rho}\right)
}{2\mathbb{R}}\right]  =0 \tag{19c}%
\end{gather}
where the quantities with a little circle "$\circ$" are defined from the
Christoffel connection (as in General Relativity). From this set eqs.(19), the
link between T and f will be determined.

ii) $f_{\mu\nu}$ has only the role to be the antisymmetric part of a
fundamental (non-symmetric) tensor K: i.e. $f_{\mu\nu}$ closed but not
necessarily exact Then, the variation of the geometrical Lagrangian
$\delta_{f}\sqrt{G}$ gives the same information that $\delta_{g}\sqrt{G}$.
that means that the remaining equations are%
\begin{gather}
R_{\left(  \mu\nu\right)  }=\overset{\circ}{R}_{\mu\nu}-T_{\mu\rho
}^{\ \ \ \alpha}T_{\alpha\nu}^{\ \ \ \rho}=-\lambda g_{\mu\nu}\tag{20a}\\
R_{\left[  \mu\nu\right]  }=\left(  \nabla_{\alpha}+2T_{\alpha}\right)
\left(  T_{\mu\nu}^{\alpha}+T_{\nu}\delta_{\mu}^{\alpha}-T_{\mu}\delta_{\nu
}^{\alpha}\right)  =-\lambda f_{\mu\nu} \tag{20b}%
\end{gather}

\subsection{Analysis and reduction of the dynamical equations}

One important equation, that appears into the two sets recently described
(independently on the specific role of the antisymmetric tensor $f_{\mu\nu}$,
bring us a lot of information about the link between $T$ and $f$ are (19b) and
(20b). Precisely, this equation $R_{\left[  \mu\nu\right]  }=-\lambda
f_{\mu\nu}$ plus the condition $\nabla_{\alpha}T_{\ \ \mu\nu}^{\alpha}=0$ lead
immediately
\begin{equation}
\nabla_{\mu}T_{\nu}-\nabla_{\nu}T_{\mu}=-\left(  \lambda f_{\mu\nu}%
+2T_{\alpha}T_{\mu\nu}^{\alpha}\right)  \tag{21}%
\end{equation}
then, the quantity that \textit{naturally appears} in the RHS is the
"definition"in the current literature of the \textit{minimal coupling
}electromagnetic tensor $\mathcal{F}_{\mu\nu}$ in an space-time with torsion.
Notice the important fact that $\nabla_{\alpha}T_{\ \ \mu\nu}^{\alpha}=0$ is
equivalent to%
\[
d^{\ast}T=0
\]
the torsion is current free. Two cases naturally arise:

i) if we assume the existence of the potential vector we have
\begin{equation}
\nabla_{\mu}T_{\nu}-\nabla_{\nu}T_{\mu}\equiv\mathcal{F}_{\mu\nu}%
=-\lambda\left(  \overset{f_{\mu\nu}}{\overbrace{\partial_{\mu}a_{\nu
}-\partial_{\nu}a_{\mu}}}\right)  -2T_{\alpha}T_{\mu\nu}^{\alpha} \tag{22}%
\end{equation}
a link between $a_{\nu}$ and $T_{\nu}$ clearly appears$:$ $T_{\nu}=-\lambda
a_{\nu}$ The important fact to remark here is that, although in references
[11] the link between the trace of the torsion and the vector potential of the
electromagnetic field was proposed, but in the theory presented in this paper
this relation is derived automatically from its geometrical basis. Beside this
point, is notable the suggestive aspect of $\mathcal{F}_{\mu\nu}$ as
$F_{\mu\nu}+B_{\mu\nu}$ with $B_{\mu\nu}$ such type of "background" field
generated by the spacetime torsion.

ii) if $f_{\mu\nu}$ has only the role to be the antisymmetric part of a
fundamental (non-symmetric) tensor K, it acquires a potential automatically,
being of this manner \textit{an exact form}$\overset{}{\text{ were }T_{\nu}}$
takes the role of potential vector. Clearly, now f cannot be potential for the
torsion from this point of view (in a non-trivial topology, it can be, of course).

From above statements over the "trace" of the torsion, is clearly seen that
two ansatz appear as candidates for the torsion tensor structure: the
"tratorial" structure $T_{\mu\nu}^{\alpha}\sim\left(  \delta_{\mu}^{\alpha
}a_{\nu}-\delta_{\nu}^{\alpha}a_{\mu}\right)  ;$ and the "product" structure
$T_{\mu\nu}^{\alpha}=k^{\alpha}f_{\mu\nu}$where the vector $k^{\alpha}$ is
eigenvector of the antisymmetric tensor $f_{\mu\nu}$, in general (notice that
torsion tensor with this "product structure" also has the possibility to be
fully antisymmetric)$.$

The other possibility is to take $\nabla_{\alpha}T_{\ \ \mu\nu}^{\alpha
}=-\lambda f_{\mu\nu}$ then $\nabla_{\mu}T_{\nu}-\nabla_{\nu}T_{\mu
}=-2T_{\alpha}T_{\mu\nu}^{\alpha},$ but their interpretation are not so clean
as before. Even more, probably carry us to a "product structure" with the
torsion tensor not fully antisymmetric, of course.

\subsection{A potential for the torsion}

As was shown in[1], if we impose the restriction $T_{\alpha\beta\gamma
}=T_{\left[  \alpha\beta\gamma\right]  }$ (e.g. totally antisymmetric torsion
tensor), from eq.(2) for example, we note that only the antisymmetric part of
the fundamental tensor $K_{\alpha\beta}$ determines fully the torsion tensor .
Then, due the assumption of a torsion tensor completely antisymmetric, the
potential torsion $f_{\mu\nu}$ exists and arises in a natural form (the
$\nabla$ for the covariant derivative with respect the full connection
$\Gamma)$. This potential torsion has the following properties
\[
f_{\mu\nu}=\overline{f}_{\mu\nu}=-f_{\nu\mu}\in\mathbb{HC}%
\]%
\begin{align}
\nabla_{\left[  \rho\right.  }f_{\left.  \mu\nu\right]  }  &  =T_{\mu\nu\rho
}\tag{23}\\
&  =\varepsilon_{\mu\nu\rho\sigma}h^{\sigma}\nonumber
\end{align}
where the last equality coming from the full antisymmetry of the Torsion
field. Immediately we can see, as a consequence of the above statements, the following\ 

i) the torsion is the dual of an axial vector $h^{\sigma}$

ii) from i), the existence in the spacetime of a completely antisymmetric
tensor covariantly constant $\varepsilon_{\mu\nu\rho\sigma}\left(
\nabla\varepsilon=0.\right)  $

Notice that, the choice for the real nature of the metric and the pure
hypercomplex potential tensor coming from the Hermitian nature of the theory:
as was clearly explained in [1].

The variational equations (in the Palatini's sense[10,12], see eqs. (12) and
(13) of ref.[1]), despite their simplest and compact form, it is necessary to
shown what is the deep physical and geometrical meaning inside these eqs..

For expression (13) of ref.[1] we have a highly nonlinear dynamical
(propagating ) equation for the torsion field, where the variation was
performed with respect to their potential $f_{\mu\nu}$ and having a nonlinear
term proportional to $f_{\mu\nu}$ playing the role of current for the
$\mathbb{T}^{\rho\sigma\tau}$. Then, the potential two form is associated
nonlinearly to the torsion field as his source regarding similar association
between the electromagnetic field and the spin in particle physics.

For the expression (12) of ref.[1], firstly is useful to split the equation
into the symmetric and the antisymmetric parts using $R_{\mu\nu}$ explicitly
as before
\begin{equation}
R_{\left(  \mu\nu\right)  }=\overset{\circ}{R}_{\mu\nu}-T_{\mu\rho
}^{\ \ \ \alpha}T_{\alpha\nu}^{\ \ \ \rho}=-2\lambda g_{\mu\nu} \tag{24}%
\end{equation}%
\begin{align}
R_{\left[  \mu\nu\right]  }  &  =\overset{\circ}{\nabla}_{\alpha}T_{\ \ \mu
\nu}^{\alpha}=-2\lambda f_{\mu\nu}\tag{25}\\
&  =\nabla_{\alpha}T_{\ \ \mu\nu}^{\alpha}\nonumber
\end{align}

(the last equality coming from the totally antisymmetry of the torsion).

Notice the important fact that $-2\lambda f_{\mu\nu}$ is the "current" for the
torsion field as the terms proportional to the 1-form potential vector
$a_{\mu}$ acts as current of the electromagnetic field $f_{\mu\nu}$ in the
equation of motion for the electromagnetic field into the standard
theory:$\nabla_{\alpha}f_{\ \ \mu}^{\alpha}=J_{\mu}$ (constants absorbed into
the $J_{\mu}$)

The symmetric part (24) can be written in a "GR" suggestive fashion%
\begin{equation}
\overset{\circ}{R}_{\mu\nu}=-2\lambda g_{\mu\nu}+T_{\mu\rho}^{\ \ \ \alpha
}T_{\alpha\nu}^{\ \ \ \rho} \tag{26}%
\end{equation}
we can advertise that the equation has the aspect of the Einstein equations
with the cosmological term modified by the torsion symmetric term $T_{\mu\rho
}^{\ \ \ \alpha}T_{\alpha\nu}^{\ \ \ \rho}$. This can be interpreted , as was
shown in [1], by the energy of the gravitational field itself.

The second antisymmetric part (25) is more involved. In order to understand
it, will be necessary use the language of differential forms to rewrite they
that, beside their symbolic and conceptual simplicity, permit us to check
consistency and covariance step by step.%
\begin{align}
\nabla_{\alpha}T_{\ \ \mu\nu}^{\alpha}  &  =-2\lambda f_{\mu\nu}\tag{27}\\
d^{\ast}T  &  =-2\lambda^{\ast}f\nonumber
\end{align}
now, using $T=\ ^{\ast}h$%
\begin{equation}
dh=-2\lambda^{\ast}f\Rightarrow\ ^{\ast}f=-\frac{1}{2\lambda}\ dh \tag{28}%
\end{equation}
in more familiar form%
\begin{equation}
\nabla_{\mu}h_{\nu}-\nabla_{\nu}h_{\mu}=-2\lambda\ ^{\ast}f_{\mu\nu} \tag{29}%
\end{equation}
then follows using again: $T=df=\ ^{\ast}h$ and eq. (27)%
\begin{equation}
d^{\ast}f=0 \tag{30}%
\end{equation}
and fundamentally%
\begin{equation}
df=-\frac{1}{2\lambda}\ d^{\ast}dh=T=\ ^{\ast}h \tag{31}%
\end{equation}%
\begin{equation}
\ d^{\ast}dh=-2\lambda\ ^{\ast}h \tag{32}%
\end{equation}
that we can recognize the Laplace-de Rham operator that help us to write the
wave covariant equation%
\begin{align}
\ \left[  \left(  d\delta+\delta d\right)  +2\lambda\right]  ^{\ast}h  &
=0\tag{33}\\
\left(  \Delta+2\lambda\right)  ^{\ast}h  &  =0\nonumber
\end{align}
If we start with the potential is not difficult to see that equivalent
equation can be find%
\begin{equation}
\left(  \Delta+2\lambda\right)  ^{\ast}f=0 \tag{34}%
\end{equation}
Notice that equation (33) coming from (28) and is consequence of the
$Tfh$-relation$\left(  T=df=\ ^{\ast}h\right)  $ but (34) comes directly from
(27). The geometric interplay between$^{\ast}$\footnotetext[2]{$^{\ast}$In
order to be consistent with the action of the Hodge operator $\left(
\ast\right)  $, in this paragraph, we assume an even number of dimensions}
\begin{equation}%
\begin{tabular}
[c]{lllllll}
&  &  & $\!\!\!$ $\ \ \ \ $\fbox{$T$} &  &  & \\
&  & $\overset{%
{\textstyle\int}
}{\swarrow}\underset{d}{\nearrow}$ &  & $\underset{_{\left(  -1\right)
^{d+1}\ast}}{\searrow}\overset{\ast}{\nwarrow}$ &  & \\
& $\ \ \ \ \ \ \ $\fbox{$\ f$} &  & $\underset{_{\overleftarrow
{\overrightarrow{-2\lambda\int^{\ast}}}}}{_{-1^{^{\ast}}d/2\lambda}}$ &  &
\fbox{$h$} &
\end{tabular}
\ \ \ \ \ \tag{35}%
\end{equation}

\section{Exact solutions in the New UFT theory}

The main motivation in this Section is clear: we must equip our "theoretical
arena" by studying wormhole solutions beyond to Einstein equations coupled to
possible matter fields. We know the that many problems appear in the
conventional "dualistic" approach even at at the classical level, that make
that the "dream" of a quantum formulation of the gravity that permit its
interaction with other fields becomes practically impossible. Then, let us
construct wormhole solutions in the viewpoint of the UFT\ model introduced
here. The action in four dimensions is given by
\begin{equation}
S=-\frac{1}{16\pi G}\int d^{4}x\sqrt{det\left\vert G_{\mu\nu}\right\vert }
\tag{36}%
\end{equation}%
\begin{equation}
\mathbb{R\equiv}\sqrt{\gamma^{4}-\frac{\gamma^{2}}{2}\overline{G}^{2}%
-\frac{\gamma}{3}\overline{G}^{3}+\frac{1}{8}\left(  \overline{G}^{2}\right)
^{2}-\frac{1}{4}\overline{G}^{4}} \tag{37}%
\end{equation}

\subsection{Totally antisymmetric torsion}

Scalar curvature $R$ and the torsion 2-form field $T_{\mu\nu}^{a}$ with a
$SU\left(  2\right)  -$Yang-Mills structure are defined in terms of the affine
connection $\Gamma_{\mu\nu\text{ }}^{\lambda}$ and the SU(2) potential torsion
$f_{\ \mu\text{ }}^{a}$by
\begin{equation}
R=g^{\mu\nu}R_{\mu\nu}\hspace{1cm}R_{\mu\nu}=R_{\mu\lambda\nu}^{\lambda}
\tag{38}%
\end{equation}%
\[
R_{\mu\lambda\nu}^{\lambda}=\partial_{\nu}\Gamma_{\mu\rho\text{ }}^{\lambda
}-\partial_{\rho}\Gamma_{\mu\nu\text{ }}^{\lambda}+...
\]%
\[
T_{\ \mu\nu}^{a}=\partial_{\mu}f_{\ \nu\text{ }}^{a}-\partial_{\nu}%
f_{\ \mu\text{ }}^{a}+\varepsilon_{bc}^{a}f_{\ \mu\text{ }}^{b}f_{\ \nu\text{
}}^{c}%
\]
$G$ and $\Lambda$ are the Newton gravitational constant and the cosmological
constant respectively. Notice the important fact that from the last equation
for the Torsion 2-form, the potential $f_{\ \mu\text{ }}^{a}$ must be
proportional with the antisymmetric part of the affine connection $\Gamma
_{\mu\nu\text{ }}^{\lambda}$ as in the Strauss-Einstein UFT. As in the case of
Einstein-Yang -Mills systems, for our new UFT\ model it can be interpreted as
a prototype of gauge theories interacting with gravity (e.g. QCD, GUTs, etc.).
Upon varying the action, we obtain the gravitational "Einstein-Eddington-like"equation%

\begin{equation}
R_{\mu\nu}=-2\lambda\left(  g_{\mu\nu}+f_{\mu\nu}\right)  \tag{39}%
\end{equation}
and the field equation for the torsion two form in differential form
\begin{equation}
d^{\ast}\mathbb{T}^{a}+\frac{1}{2}\varepsilon^{abc}\left(  f_{b}\wedge^{\ast
}\mathbb{T}_{c}-^{\ast}\mathbb{T}_{b}\wedge f_{c}\right)  =\mathbb{F}^{a}
\tag{40}%
\end{equation}
where we define as usual
\[
\mathbb{T}_{\ bc}^{a}\equiv\frac{\partial L_{G}}{\partial T_{a}^{\ bc}%
},\mathbb{F}_{\ bc}^{a}\equiv\frac{\partial L_{G}}{\partial F_{a}^{\ }}%
\]
we are going to seek for a classical solution of eqs. (39) and (40) with the
following spherically symmetric ansatz for the metric and gauge connection
\begin{equation}
ds^{2}=d\tau^{2}+a^{2}\left(  \tau\right)  \sigma^{i}\otimes\sigma^{i}\equiv
d\tau^{2}+e^{i}\otimes e^{i} \tag{41}%
\end{equation}
here $\tau$ is the euclidean time and the dreibein is defined by $e^{i}\equiv
a\left(  \tau\right)  \sigma^{i}.$The gauge connection is
\begin{equation}
f^{a}\equiv f_{\mu}^{a}dx^{\mu}=h\sigma^{a} \tag{42}%
\end{equation}
for $a=1,2,3$ and for $a=0$%

\begin{equation}
f^{0}\equiv f_{\mu}^{0}dx^{\mu}=s\sigma^{0} \tag{43}%
\end{equation}
this choice for the potential torsion is the most general and consistent from
the physical and mathematical point of view due the symmetries involved in the
problem, as we will show soon.

The $\sigma^{i}$ one-form satisfies the $SU\left(  2\right)  $ Maurer-Cartan
structure equation
\begin{equation}
d\sigma^{a}+\varepsilon_{\ bc}^{a}\sigma^{b}\wedge\sigma^{c}=0 \tag{44}%
\end{equation}
Notice that in the ansatz the frame and isospin indexes are identified as for
the case with the NBI Lagrangian of ref.[3]. The torsion two-form
\begin{equation}
T^{\gamma}=\frac{1}{2}T_{\ \mu\nu}^{\gamma}dx^{\mu}\wedge dx^{\nu} \tag{45}%
\end{equation}
becomes
\begin{align}
T^{a}  &  =df^{a}+\frac{1}{2}\varepsilon_{\ bc}^{a}f^{b}\wedge f^{c}\tag{46}\\
&  =\left(  -h+\frac{1}{2}h^{2}\right)  \varepsilon_{\ bc}^{a}\sigma^{b}%
\wedge\sigma^{c}\nonumber
\end{align}
Notice that $f^{0}$ plays no role here because we take simply $\ ds=0$ (the
$U\left(  1\right)  $ component of $SU\left(  2\right)  ,$ in principle, does
not form part of the space spherical symmetry) , and the expression for the
torsion is analogous to the non abelian two form strength field of [3]. Is
important to note that, when we goes from the Lorentzian to Euclidean
gravitational regime, $it\rightarrow\tau$ and the torsion pass from the field
of the \textit{Hypercomplex} to the \textit{Complex} numbers, for invariance
reasons (geometrically, multiplication of hypercomplex numbers preserves the
(square) Minkowski norm $(x^{2}-y^{2})$ in the same way that multiplication of
complex numbers preserves the (square) Euclidean norm $(x^{2}+y^{2})$).
Inserting $T^{a}$ from eq. (46) into the dynamical equation (40) we obtain
\begin{equation}%
\begin{array}
[c]{l}%
d^{\ast}\mathbb{T}^{a}+\frac{1}{2}\varepsilon^{abc}\left(  f_{b}\wedge^{\ast
}\mathbb{T}_{c}-^{\ast}\mathbb{T}_{b}\wedge f_{c}\right)  =\mathbb{\ }^{\ast
}\mathbb{F}^{a}\\
(-2h+h^{2})(1-h)d\tau\wedge e^{b}\wedge e^{c}=-2\lambda d\tau\wedge
e^{b}\wedge e^{c}%
\end{array}
\tag{47}%
\end{equation}
where
\begin{equation}
^{\ast}\mathbb{T}^{a}\mathbb{\equiv}\frac{\lambda\sqrt{\left\vert g\right\vert
}}{\sqrt{3}}hA(-2h+h^{2})d\tau\wedge\frac{e^{a}}{a^{2}} \tag{48}%
\end{equation}%
\begin{equation}
^{\ast}\mathbb{F}^{a}=-\frac{2\lambda^{2}\sqrt{\left\vert g\right\vert }%
}{\sqrt{3}}hA\frac{d\tau\wedge e^{b}\wedge e^{c}}{a^{3}} \tag{49}%
\end{equation}%
\begin{equation}
A\equiv\lambda^{4}\left[  \left(  1+\alpha\right)  ^{2}+\alpha/2\right]  ,
\tag{50}%
\end{equation}
and%
\begin{equation}
\alpha=\frac{1}{2}\left(  s^{2}+3h^{2}\right)  , \tag{51}%
\end{equation}
from expression (47) we have an algebraic cubic equation for $h$%
\begin{equation}
(-2h+h^{2})(1-h)+2\lambda=0 \tag{52}%
\end{equation}

We can see that, in contrast with our previous work with a dualistic theory
[3] where the energy-momentum tensor of Born-Infeld was considered, for $h$
there exist three non trivial solutions depending on the cosmological constant
$\lambda.$ But, at this preliminary analysis of the problem, only the values
of h that make the quantity$\left(  -h+\frac{1}{2}h^{2}\right)  $
$\in\mathbb{R}$ are relevant for our proposes: due the pure imaginary
character of $T$ in the euclidean framework and mainly to compare with the
NABI wormhole solution of our previous work (the question of the
h$\in\mathbb{C}$ will be the focus of a further paper [5]). As the value of
$h$ $\in\mathbb{R}$ is -1 and in 4 spacetime dimensions $\lambda=\left\vert
1-d\right\vert =3,$ then
\begin{equation}
\left.  T_{bc}^{a}\right\vert _{h_{1}}=\frac{3}{2}\frac{\varepsilon_{bc}^{a}%
}{a^{2}};\ \ \ \ \ \ \ \hspace{1.54cm}T_{0c}^{a}=0 \tag{53}%
\end{equation}
Namely, only the magnetic field is non vanishing while the electric field
vanishes. An analogous feature can be seen in the solution of Giddings and
Strominger and in our previous paper[3]. Substituting the expression for the
Torsion two form (53) into the symmetric part of the variational equation,
namely$^{1}$\footnotetext[1]{in the tetrad: $\overset{\circ}{R}_{_{00}%
}=-3\frac{\overset{\cdot\cdot}{a}}{a},\overset{\circ}{R}_{ab}=-\left[
\frac{\overset{\cdot\cdot}{a}}{a}+2\left(  \frac{\overset{\cdot}{a}}%
{a}\right)  ^{2}-\frac{2}{a^{2}}\right]  $}%
\begin{equation}
R_{\left(  \mu\nu\right)  }=\overset{\circ}{R}_{\mu\nu}-T_{\mu\rho
}^{\ \ \ \alpha}T_{\alpha\nu}^{\ \ \ \rho}=-2\lambda g_{\mu\nu} \tag{54}%
\end{equation}
we reduce the equation (24) to an ordinary differential equation for the scale
factor $a$,%
\begin{align}
\left[  \left(  \frac{\overset{.}{a}}{a}\right)  ^{2}-\frac{1}{a^{2}}\right]
&  =\frac{2\lambda}{3}-\frac{9}{2a^{4}},\ \ \ \ \ \tag{56}\\
\frac{Ln\left[  1+4a^{2}+2\sqrt{-9+2a^{2}+4a^{4}}\right]  }{2\sqrt{2}}  &
=\tau-\tau_{0} \tag{54}%
\end{align}%
\begin{gather}
\ T_{\mu\rho}^{\ \ \ \alpha}T_{\alpha\nu}^{\ \ \ \rho}=\frac{\left(
-h+\frac{1}{2}h^{2}\right)  ^{2}}{a^{4}}2\delta_{\mu\nu}\tag{57}\\
=\frac{9}{2a^{4}}\delta_{\mu\nu}\nonumber
\end{gather}
There are 2 values for the scale factor $a$: max. and min. respectively,
namely%
\begin{equation}
a=\mp\frac{e^{-\sqrt{2}\left(  \tau-\tau_{0}\right)  }\sqrt{37-2e^{2\sqrt
{2}\left(  \tau-\tau_{0}\right)  }+e^{4\sqrt{2}\left(  \tau-\tau_{0}\right)
}}}{2\sqrt{2}} \tag{58}%
\end{equation}
Expression (58) for the scale factor $a$ is described in the Figure 1 for the
real value of $h$.

As is easily seen from (58), the scale factor has an exponentially growing
behavior, in sharp contrast to the wormhole solution from our previous work
with the "dualistic" non-abelian BI\ theory Figure 4. Also, for this
particular value of the torsion, the wormhole tunneling interpretation (in the
sense of the Coleman' s mechanism) is fulfilled. Now will need to see what
happens with the equation (27) in this particular case under consideration:
equation (27) takes the following form
\begin{equation}%
\begin{array}
[c]{l}%
d^{\ast}T^{a}+\frac{1}{2}\varepsilon^{abc}\left(  f_{b}\wedge^{\ast}%
T_{c}-^{\ast}T_{b}\wedge f_{c}\right)  =-2\lambda\mathbb{\ }^{\ast}f^{a}\\
(-2h+h^{2})(1-h)d\tau\wedge e^{b}\wedge e^{c}=-2\lambda d\tau\wedge
e^{b}\wedge e^{c}%
\end{array}
\tag{59}%
\end{equation}%
\begin{equation}
^{\ast}T^{a}\mathbb{\equiv}h(-2h+h^{2})d\tau\wedge\frac{e^{a}}{a^{2}} \tag{60}%
\end{equation}%
\begin{equation}
^{\ast}f^{a}=-h\frac{d\tau\wedge e^{b}\wedge e^{c}}{a^{3}} \tag{61}%
\end{equation}
Then we arrived to the same equation for $\lambda$ as (52) corroborating the
self-consistency of the procedure.

\subsection{ "Tratorial" torsion}

For begin with, let us consider the problem involving the set of eq. (19) with
the usual definition for the SU(2) electromagnetic field strength%

\begin{equation}
f^{\gamma}=\frac{1}{2}f_{\ \mu\nu}^{\gamma}dx^{\mu}\wedge dx^{\nu} \tag{62}%
\end{equation}
and as before, we are going to seek for a classical solution of eqs. (19) with
the following spherically symmetric ansatz for the metric and gauge
connection
\begin{equation}
ds^{2}=d\tau^{2}+a^{2}\left(  \tau\right)  \sigma^{i}\otimes\sigma^{i}\equiv
d\tau^{2}+e^{i}\otimes e^{i} \tag{63}%
\end{equation}
here $\tau$ is the euclidean time and the dreibein is defined by $e^{i}\equiv
a\left(  \tau\right)  \sigma^{i}.$ However, in the case of the set (19) we
have been assume that the two form $f^{\gamma}$ comes from a 1-form potential
$A$ where, as in the non abelian Born-Infeld model of ref.[3], is defined as
$A^{a}\equiv A_{\mu}^{a}dx^{\mu}=h\sigma^{a}$.

The extremely important fact in this case is that we know that $\sigma^{i}$
one-form satisfies the $SU\left(  2\right)  $ Maurer-Cartan structure
equation, as fundamental geometrical structure of the non-abelian
electromagnetic field%
\begin{equation}
d_{su(2)}\sigma^{a}+\varepsilon_{\ bc}^{a}\sigma^{b}\wedge\sigma^{c}=0
\tag{64}%
\end{equation}
but now due the identification assumed in (63):
\begin{align}
e^{i}  &  \equiv a\left(  \tau\right)  \sigma^{i}.\tag{65}\\
&  \Rightarrow de^{a}=T^{a}-e_{\ b}^{a}\wedge\sigma^{b} \tag{66}%
\end{align}
here we make the difference between the exterior derivatives in the spacetime
with torsion and in the SU(2) group manifold. Is clearly seen that a question
of compatibility involving the identification of the gauge group with the
geometrical structure of the space-time with torsion certainly exists. From
(64-66) we see that%
\begin{equation}
\partial_{\tau}ad\tau\wedge\sigma^{a}-a\varepsilon_{\ bc}^{a}\sigma^{b}%
\wedge\sigma^{c}=T^{a}-e_{\ b}^{a}\wedge\sigma^{b} \tag{67}%
\end{equation}
If
\begin{equation}
e_{\ b}^{a}=-\varepsilon_{\ bc}^{a}\sigma^{c} \tag{68}%
\end{equation}
and%
\begin{equation}
T^{a}=\delta_{b}^{a}\left(  \partial_{\tau}a\right)  d\tau\wedge\sigma^{b}
\tag{69}%
\end{equation}
the space-time and gauge group are fully compatible then%
\begin{equation}
d\sigma^{a}+\varepsilon_{\ bc}^{a}\sigma^{b}\wedge\sigma^{c}=0 \tag{70}%
\end{equation}
is restored. Hence, the general form assumed for the torsion field, due the
symmetry conditions prescribed above, is%
\begin{equation}
T_{\beta\gamma}^{\alpha}=\xi\left(  \delta_{\beta}^{\alpha}u_{\gamma}%
-\delta_{\gamma}^{\alpha}u_{\beta}\right)  +\varsigma h_{\delta}%
\varepsilon^{\delta\alpha}{}_{\beta\gamma}\ \ \ \ \left(  \xi,\varsigma
:const.\right)  \tag{71}%
\end{equation}
Notice that the condition of compatibility that impose such type of "trator"
form for the torsion tensor in order to restore the behaviour of the volume
form of the space-time with respect to the covariant derivative, here appear
in a natural manner without introduce any extra scalar field (dilaton) or to
pass to other frame (i.e.: Jordan, Einstein,etc.). Moreover, if we have been
continue without make the correspondences (68-69), the equations of motion for
the electromagnetic field itself bring automatically these conditions (see in
the next paragraph).

Notice that in the HO ansatz the frame and isospin indexes are identified as
for the case with the NBI Lagrangian of ref.[3]. The electromagnetic field
two-form
\begin{align}
f^{a}  &  =dA^{a}+\frac{1}{2}\varepsilon_{\ bc}^{a}A^{b}\wedge A^{c}\tag{72}\\
&  =h\delta_{b}^{a}\left(  \partial_{\tau}\ln a\right)  d\tau\wedge\sigma
^{b}+h\frac{T^{a}}{a}-\left(  -h+\frac{1}{2}h^{2}\right)  \varepsilon
_{\ bc}^{a}\sigma^{b}\wedge\sigma^{c}\nonumber\\
&  =\left(  -h+\frac{1}{2}h^{2}\right)  \varepsilon_{\ bc}^{a}\sigma^{b}%
\wedge\sigma^{c}\nonumber
\end{align}
where in the last equality conditions (68-69) have been assumed. The dynamical
eqs.%
\[
\mathbb{F}_{\ bc}^{a}\equiv\frac{\partial L_{G}}{\partial F_{a}^{\ }%
}\Rightarrow
\]%
\begin{equation}
^{\ast}\mathbb{F}^{a}\mathbb{\equiv}\frac{\lambda\sqrt{\left\vert g\right\vert
}}{\sqrt{3}}h\mathbb{A}(-2h+h^{2})d\tau\wedge\frac{e^{a}}{a^{2}}\equiv
Mh(-2h+h^{2})d\tau\wedge\frac{e^{a}}{a^{2}} \tag{73}%
\end{equation}
Inserting it in the Yang-Mills type field equation (19c) we obtain
\begin{equation}%
\begin{array}
[c]{l}%
d^{\ast}\mathbb{F}^{a}+\frac{1}{2}\varepsilon^{abc}\left(  A_{b}\wedge^{\ast
}\mathbb{F}_{c}-^{\ast}\mathbb{F}_{b}\wedge A_{c}\right)  =0\\
=Mh\ d\tau\wedge\sigma^{b}\wedge\sigma^{c}\left(  -2h+h^{2}\right)  \left(
h-1\right)
\end{array}
\tag{74}%
\end{equation}%
\[
\mathbb{A}\equiv\lambda^{4}\left[  \left(  1+\alpha\right)  ^{2}%
+\alpha/2\right]
\]
Then, there exists a non trivial solution: h=1,$\left(  \text{with s=0 in
}\mathbb{A}\text{ as before in[1].}\right)  .$ The electromagnetic field is
immediately determined, and is as in the non abelian Born-Infeld model of our
previous reference and in the result of Giddings and Strominger, namely%
\begin{equation}
f_{\ bc}^{a}=-\frac{\varepsilon_{\ bc}^{a}}{a^{2}}\ \ \ \ \ \ \ \ \ \ f_{\ 0c}%
^{a}=0 \tag{75}%
\end{equation}

only we have magnetic field.

Now considering only a "trator" form for the torsion, eq.(16b) is identically
null due the magnetic character of $f^{a}$ and the particular form of the
symmetric coefficients of the connection. Inserting the torsion eq.(69) into
the eq. (19a), as in previous section, we obtain%
\begin{equation}
\left[  \left(  \frac{\overset{.}{a}}{a}\right)  ^{2}-\frac{1}{a^{2}}\right]
=\frac{\lambda}{3} \tag{76}%
\end{equation}
Integration of this last expression immediately leads%
\begin{equation}
a\left(  \tau\right)  =\left(  \frac{\lambda}{3}\right)  ^{-1/2}\ Sinh\left[
\left(  \frac{\lambda}{3}\right)  ^{1/2}\ \left(  \tau-\tau_{0}\right)
\right]  \tag{77}%
\end{equation}
Then is quite evident that this particular case doesn't lead wormhole
configurations: only eternal expansion with $a\left(  \tau_{0}\right)  =0$
(the origin of the euclidean time Fig.2).

Now considering only the product form for the torsion, eq. (19c) doesn't
change but eq.(19b) takes the form of a wave equation for the scale factor
\[
\left[  \square a+\left(  \partial_{0}a\right)  \left(  \partial^{0}a\right)
\right]  =\lambda
\]
due $T_{\beta\gamma}^{\alpha}=\varsigma k^{\alpha}\varepsilon_{\beta\gamma
}\rightarrow\varepsilon_{ab}\left(  \partial^{0}a\right)  $ Is not difficult
to see that the su(2) structure of the electromagnetic tensor is of some
manner transferred to the structure of the torsion. But here we enter in
conflict because the system of eqs. (19) turns to be\textit{ overdetermined}:
probably we need more freedom in the ansatz for $f_{bc}^{a}$ (s$\neq0,$ or
h=h$\left(  \tau\right)  )$. This fact will be studied in near future [5].

\subsection{General case}

Let to assume the full form (71) for $T^{a}$%
\begin{align}
^{\ast}\mathbb{F}^{a}  &  \equiv Mh\left\{  h\delta_{d}^{a}\left(
\partial_{\tau}\ln a\right)  \varepsilon_{\ \ \ ed}^{d0}\sigma^{e}\wedge
\sigma^{d}+\right. \tag{78}\\
&  +\frac{h}{a}\left[  \xi\left(  \delta_{i}^{a}u_{j}-\delta_{j}^{a}%
u_{i}\right)  +\varsigma h_{\delta}\varepsilon^{\delta a}{}_{ji}\right]
\varepsilon_{\ \ \ kl}^{ij}\omega^{k}\wedge\omega^{l}+\nonumber\\
&  \left.  +(-2h+h^{2})\varepsilon_{\ bc}^{a}\varepsilon_{\ \ \ 0d}^{bc}%
d\tau\wedge\sigma^{a}\right\} \nonumber
\end{align}
here, in order to avoid the cumbersome expression in the second term due the
standard orthonormal splitting, $ij=0,a,b,c$ and the $\omega^{k}$ are the
corresponding 1-forms ($d\tau,\sigma^{a}..)$ wherever the case. The YM type
equation can be written as%
\begin{gather}
d^{\ast}\mathbb{F}^{a}+\frac{1}{2}\varepsilon^{abc}\left(  A_{b}\wedge^{\ast
}\mathbb{F}_{c}-^{\ast}\mathbb{F}_{b}\wedge A_{c}\right)  =\tag{79}\\
Mh\left\{  \left[  h\delta_{b}^{a}\left(  \partial_{\tau}\partial_{\tau}\ln
a\right)  +\partial_{\tau}\left(  \frac{h}{a}\left(  \xi\left(  \delta_{b}%
^{a}u_{0}-\delta_{0}^{a}u_{b}\right)  +\varsigma h_{c}\varepsilon^{ca}{}%
_{b0}\right)  \right)  \right]  \varepsilon_{\ \ \ ed}^{b0}d\tau\wedge
\sigma^{e}\wedge\sigma^{d}+\right. \nonumber\\
\left.  \left[  h\delta_{b}^{a}\left(  \partial_{\tau}\ln a\right)  +\frac
{h}{a}\left(  \xi\left(  \delta_{b}^{a}u_{0}-\delta_{0}^{a}u_{b}\right)
+\varsigma h_{c}\varepsilon^{ca}{}_{b0}\right)  \right]  2d\left(  \sigma
^{e}\wedge\sigma^{d}\right)  \right\}  +\nonumber\\
+M\left[  \frac{h}{a}\left(  \xi\left(  \delta_{b}^{a}u_{0}-\delta_{0}%
^{a}u_{b}\right)  +\varsigma h_{c}\varepsilon^{ca}{}_{b0}\right)
+(-2h+h^{2})\right]  (h-1)d\tau\wedge\sigma^{b}\wedge\sigma^{c}=0\nonumber
\end{gather}
from the above equation we obtain information about the determination of the
$f$ field and of the torsion field as in the previous cases: the first term%
\begin{equation}
\left[  h\delta_{b}^{a}\left(  \partial_{\tau}\partial_{\tau}\ln a\right)
+\partial_{\tau}\left(  \frac{h}{a}\left(  \xi\left(  \delta_{b}^{a}%
u_{0}-\delta_{0}^{a}u_{b}\right)  +\varsigma h_{c}\varepsilon^{ca}{}%
_{b0}\right)  \right)  \right]  =0 \tag{80}%
\end{equation}
leads immediately%
\begin{gather}
\left[  \eta_{ab}\partial_{0}a+\left(  \xi\left(  \eta_{ab}u_{0}-\eta
_{a0}u_{b}\right)  +\varsigma h_{c}\varepsilon^{ca}{}_{b0}\right)  \right]
=\Xi_{ab0}^{A}+\Xi_{ab0}^{s}\tag{81}\\
\Rightarrow\varsigma h_{c}\varepsilon^{c}{}_{ab0}\equiv\Xi_{ab0}%
^{A}\nonumber\\
\Rightarrow\eta_{ab}\partial_{0}a+\xi\left(  \eta_{ab}u_{0}-\eta_{a0}%
u_{b}\right)  =\Xi_{ab0}^{S}\nonumber
\end{gather}
where the tensor
\[
\Xi_{ab0}=\Xi_{ab0}^{A}+\Xi_{ab0}^{S}%
\]
is independent of the time, and the superscripts $A$ and $S$ indicate the
totally antisymmetric part of the another non-totally antisymmetric. Then, the
second and third equalities above follows. Is not difficult to see, that
contracting indices, tracing and considering the symmetries involved, we
obtain explicitly%
\begin{align}
T_{b0}^{a}  &  =\delta_{\left[  b\right.  }^{a}\partial_{\left.  0\right]
}a-a\widetilde{\Xi}_{\ \ \ b0}^{Sa}+\Xi_{ab0}^{A}\tag{82}\\
T_{bc}^{a}  &  =-a\widetilde{\Xi}_{\ \ \ bc}^{Sa}+\varsigma h_{0}%
\varepsilon^{0a}{}_{bc}\tag{83}\\
T_{bc}^{0}  &  =-a\widetilde{\Xi}_{\ \ \ bc}^{S0}+\varsigma h_{c}%
\varepsilon^{c0}{}_{bc} \tag{84}%
\end{align}
where the integration tensor (independent on time) are related with $u_{i}$
and $\widetilde{\Xi}_{\ \ \ kl}^{Sj}(ij..=0,a,b,c)$ as follows:
\[
u_{c}=-\frac{a\Xi_{c}^{S}}{2\xi},u_{0}=-\frac{1}{2\xi}\left(  3\partial
_{0}a+a\Xi_{c}^{S}\right)  ,\Xi_{c}^{S}\equiv\Xi_{\ \ cj}^{Sj},\Xi_{0}%
^{S}\equiv\Xi_{\ \ 0j}^{Sj}and\ \widetilde{\Xi}_{\ \ \ kl}^{Sj}\equiv\frac
{-1}{2}\left(  \delta_{k}^{j}\Xi_{l}^{S}-\delta_{l}^{j}\Xi_{k}^{S}\right)
\]
The last term, however, indicate us that there exist a simplest solution with
$h=1,$ as the previous case for the non abelian $f.$ Then
\[
f_{\ bc}^{a}=-\frac{\varepsilon_{\ bc}^{a}}{a^{2}}\ ,\ f_{\ b0}^{a}=0
\]
again, and the second is identically cero due the symmetry of the torsion
2-form with respect to the tetrad defined by (63).\ Now the question is if the
system of equations is overdetermined or not: $k^{a}$ and $\ a$ are without
determine. To this end, we carry the information into the expressions (82-84)
to the second equation of the set, namely eq.(19b). Again, the symmetry
involved both : from the equations%
\begin{align}
\nabla_{i}T_{ab}^{i}+2T_{i}T_{ab}^{i}  &  =-\lambda f_{ab}^{c}e_{c}\tag{85}\\
\nabla_{i}T_{ao}^{i}+2T_{i}T_{a0}^{i}  &  =0 \tag{86}%
\end{align}
fix the torsion tensor components as%
\begin{align}
T_{b0}^{a}  &  =\delta_{\left[  b\right.  }^{a}\partial_{\left.  0\right]
}a\tag{87}\\
T_{bc}^{a}  &  =-a\widetilde{\Xi}_{\ \ \ bc}^{Sa}+\varsigma h_{0}%
\varepsilon^{0a}{}_{bc}\tag{88}\\
T_{bc}^{0}  &  =0 \tag{89}%
\end{align}
Expression (86) turns a null identity, and from (85) only%
\begin{gather}
4T_{i}T_{ab}^{i}=4a\widetilde{\Xi}_{c}^{S}\overset{T_{bc}^{a}}{\overbrace
{\left(  -a\widetilde{\Xi}_{\ \ \ bc}^{Sa}+\varsigma h_{0}\varepsilon^{0a}%
{}_{bc}\right)  }}=-\lambda f_{ab}^{c}e_{c}\tag{90}\\
\Rightarrow4\left(  a^{2}\widetilde{\Xi}_{c}^{S}\widetilde{\Xi}_{\ \ \ ab}%
^{Sc}-a\widetilde{\Xi}_{c}^{S}\varsigma h_{0}\varepsilon^{0c}\varepsilon
_{ab}\right)  =-\lambda f_{ab}^{c}e_{c}\nonumber\\
a\widetilde{\Xi}_{c}^{S}\varsigma h_{0}\varepsilon^{0c}\varepsilon
_{ab}=-\lambda f_{ab}^{c}e_{c}=\frac{\lambda\varepsilon_{ab}^{c}}{a^{2}}%
e_{c}\nonumber\\
\widetilde{\Xi}_{c}^{S}\varsigma h_{0}\varepsilon^{0c}\varepsilon_{ab}%
=\frac{\lambda\varepsilon_{ab}^{c}}{a^{2}}\sigma_{c}\nonumber
\end{gather}
where in the last line with use the property $\widetilde{\Xi}_{c}%
^{S}\widetilde{\Xi}_{\ \ \ ab}^{Sc}=\Xi_{c}^{S}\left(  \delta_{a}^{c}\Xi
_{b}^{S}-\delta_{b}^{c}\Xi_{a}^{S}\right)  \equiv0$ (see definitions above).

Is easily seen, that squaring both sides of (90) and from (89) we obtain
\[
h_{0}=\frac{\lambda\sigma_{0}}{a^{2}\left\vert \widetilde{\Xi}_{c}%
^{S}\right\vert 2\varsigma},\text{ \ \ \ \ \ \ \ \ \ \ }h_{c}=\frac
{\lambda\left\vert \widetilde{\Xi}_{c}^{S}\right\vert a\sigma_{c}}{2\varsigma
},
\]
and analogically to the previous cases, from the eqs. (19a) the equation to
integrate takes the form%
\[
\frac{da}{d\tau}=\pm\frac{4}{5}\left[  1+\frac{\lambda}{3}a^{2}+\frac{2}%
{3}a^{4}\left\vert \widetilde{\Xi}_{c}^{S}\right\vert ^{2}+\frac{3}{8}\left(
\frac{\lambda}{\left\vert \widetilde{\Xi}_{c}^{S}\right\vert a}\right)
^{2}\right]  ^{1/2}%
\]
One interesting case when the above equation can be integrated exactly is
precisely when d=4. This condition, besides improving the integrability
condition of the equation, fix $\left\vert \widetilde{\Xi}_{c}^{S}\right\vert
^{2}>3/2$. The scale factor $a\left(  \tau\right)  $ takes the following form%

\[
a\left(  \tau\right)  =\sqrt{B+(A-B)Tanh^{2}\left[  \frac{\left(  \tau
-\tau_{0}\right)  \sqrt{(A-B)}}{2}\right]  }%
\]
where A and B are nonlinear functions of the norm square $\left\vert
\widetilde{\Xi}_{c}^{S}\right\vert ^{2}.$The explicit form of these functions
are not crucial: only the bound for $\left\vert \widetilde{\Xi}_{c}%
^{S}\right\vert ^{2}>3/2$ need to be preserved\ (also through the
normalization of A and B into the graphic representation i.e. Fig. 3) Notice
that the spacetime is asymptotically Minkowskian with a throat $a\left(
\tau_{0}\right)  $ =$\sqrt{B}$ (however the values of the constants have been
selected according the previous remarks). Other possibilities not enumerated
here, lead spacetimes with cyclic singularities due trascendental functions
into the denominator of the expression for the scale factor $a\left(
\tau\right)  $. This issue is a focus of a future discussion somewhere [5].

\subsection{Coexistence of both type of torsion in cosmological spacetimes}

Is interesting to note that in reference [4] the field equations of vacuum
quadratic Poincare gauge field theory (QPGFT) were solved for purely null
tratorial torsion. The author there expressing the contortion tensor for such
a case as%
\[
K_{\lambda\mu\nu}=-2(g_{\lambda\mu}a_{\nu}-g_{\lambda\nu}a_{\mu})
\]
However, the important thing is that the author have been discussed the
relationship between this class (tratorial) and a similar class of solution
with null axial vector torsion, arriving to the conclusion that cosmological
solutions with different type of torsion are forbidden. The main reason of
this situation can have 2 origins: the specific theory and action (QPGFT), or
the Newman-Penrose method used in the computations that works, as is well
know, with null geometric quantities. Here we shown that this problem not
arises in our theory.

\section{The underlying Dirac structure of the spacetime manifold}

The real structure of the Dirac equation in the form%
\begin{align}
\left(  \gamma_{0}p_{0}-i\gamma\cdot\mathbf{p}\right)  \mathbf{u}  &
=m\mathbf{v}\tag{91}\\
\left(  \gamma_{0}p_{0}+i\gamma\cdot\mathbf{p}\right)  \mathbf{v}  &
=m\mathbf{u} \tag{92}%
\end{align}
with%
\begin{equation}
\gamma_{0}=\left(
\begin{array}
[c]{cc}%
\sigma_{0} & 0\\
0 & \sigma_{0}%
\end{array}
\right)  ,\ \ \ \ \ \gamma=\left(
\begin{array}
[c]{cc}%
0 & -\sigma\\
\sigma & 0
\end{array}
\right)  \tag{93}%
\end{equation}
where $\sigma$ are the Pauli matrices and $p=\left(  \widehat{p}_{1}%
,\widehat{p}_{2},\widehat{p}_{3}\right)  ,$ determines a 4D real vector space
with G as its automorphism, such that $G\subset L\left(  4\right)  $. This
real vector space can be make coincides with the tangent space to the
spacetime manifold $M,$ being this the idea. The principal fiber bundle (PFB)
$P\left(  G,M\right)  $ with the structural group $G$ determines the (Dirac)
geometry of the spacetime. We suppose now $G$ with the general form%
\begin{equation}
G=\left(
\begin{array}
[c]{cc}%
A & B\\
-B & A
\end{array}
\right)  ,\text{ \ \ \ }G^{+}G=I_{4} \tag{94}%
\end{equation}
A,B\ 2$\times$2 matrices. Also there exists a fundamental tensor $J_{\mu
}^{\ \ \lambda}J_{\lambda}^{\ \ \nu}=\delta_{\mu}^{\nu}$ invariant under $G$
with structure
\begin{equation}
J=\left(
\begin{array}
[c]{cc}%
0 & \sigma_{0}\\
-\sigma_{0} & 0
\end{array}
\right)  \tag{95}%
\end{equation}
where however, the Lorentz metric $g_{\lambda\mu}$ is also invariant under
$G.$due its general form (94). Finally, a third fundamental tensor
$\sigma_{\lambda\mu}$ is also invariant under $G$ where the following
relations between the fundamental tensors are%
\begin{equation}
J_{\lambda}^{\ \ \nu}=\sigma_{\lambda\mu}g^{\lambda\nu},\text{ \ \ \ \ \ \ }%
g_{\mu\nu}=\sigma_{\lambda\mu}J_{\nu}^{\text{ }\lambda},\text{ \ \ \ \ \ \ }%
\sigma_{\lambda\mu}=J_{\lambda}^{\text{ }\nu}g_{\mu\nu} \tag{96}%
\end{equation}
where%
\begin{equation}
g^{\lambda\nu}=\frac{\partial g}{\partial g_{\lambda\nu}}\text{ \ \ }\left(
g\equiv\det(g_{\mu\nu})\right)  \tag{97}%
\end{equation}
Then, the necessary fundamental structure is given by
\begin{equation}
G=L\left(  4\right)  \cap Sp\left(  4\right)  \cap K\left(  4\right)  \tag{98}%
\end{equation}
which leaves concurrently invariant the three fundamental forms
\begin{align}
ds^{2}  &  =g_{\mu\nu}dx^{\mu}dx^{\nu}\tag{99}\\
\sigma &  =\sigma_{\lambda\mu}dx^{\lambda}\wedge dx^{\mu}\tag{100}\\
\phi &  =J_{\nu}^{\text{ }\lambda}w^{\nu}v_{\lambda} \tag{101}%
\end{align}
where $w^{\nu}$ are components of a vector $w^{\nu}\in V^{\ast}:$ the dual
vector space. In expression (98) $L\left(  4\right)  $ is the Lorentz group in
4D, $Sp\left(  4\right)  $ is the Symplectic group in 4D real vector space and
$K\left(  4\right)  $ denotes the almost complex group that leaves $\phi$ invariant[6].

For instance, $G$ leaves the geometric product invariant [7]%
\begin{align}
\gamma_{\mu}\gamma_{\nu}  &  =\frac{1}{2}\left(  \gamma_{\mu}\gamma_{\nu
}-\gamma_{\nu}\gamma_{\mu}\right)  +\frac{1}{2}\left(  \gamma_{\mu}\gamma
_{\nu}+\gamma_{\nu}\gamma_{\mu}\right) \nonumber\\
&  =\gamma_{\mu}\cdot\gamma_{\nu}-\gamma_{\mu}\wedge\gamma_{\nu}=g_{\mu\nu
}+\sigma_{\mu\nu} \tag{102}%
\end{align}
where the are now regarded as a set of orthonormal basis vectors , of such a
manner that any vector can be represented as $\mathbf{v}=v^{\lambda}%
\gamma_{\lambda}$ and
\begin{equation}
\varepsilon_{\alpha\beta\gamma\delta}\equiv\gamma_{\alpha}\wedge\gamma_{\beta
}\wedge\gamma_{\gamma}\wedge\gamma_{\delta} \tag{103}%
\end{equation}

In resume, the fundamental structure of the spacetime is then represented by
P$\left(  G,M\right)  ,$ where $G$ is given by $\left(  98\right)  ,$ which
leaves invariant the fundamental forms $\left(  99-101\right)  ,$ implying
that%
\begin{align}
\nabla_{\lambda}g_{\mu\nu}  &  =0\tag{104}\\
\nabla_{\nu}\sigma_{\lambda\mu}  &  =0\tag{105}\\
\nabla_{\lambda}J_{\nu}^{\text{ }\lambda}  &  =0 \tag{106}%
\end{align}
where $\nabla_{\lambda}$ denotes the covariant derivative of the $G$
connection. Is interesting to note that it is only necessary to consider two
of above three equations: the third follows automatically. Then, we will
consider $\left(  104\right)  \left(  105\right)  $ because in some sense they
represents the boson and fermion symmetry respectively.

\subsection{Field equations and group structure}

Is necessary to introduce now other antisymmetric tensor $\sigma_{\mu\nu
}^{\prime}$ which is not helical, that means that is different of $\sigma
_{\mu\nu}$ of (102) but also invariant with respect to the generalized
connection $G:\nabla_{\nu}\sigma_{\lambda\mu}=0$. For instance, we can
construct also the antisymmetric tensor $\vartheta_{\mu\nu}\equiv\sigma
_{\mu\nu}^{\prime}-\sigma_{\lambda\mu}\neq0,$ that obeys $\nabla_{\nu
}\vartheta_{\mu\nu}=0$ and obviously $\frac{1}{6}\left(  \partial_{\mu
}\vartheta_{\nu\lambda}+\partial_{\nu}\vartheta_{\lambda\mu}+\partial
_{\lambda}\vartheta_{\mu\nu}\right)  =T_{\ \nu\mu}^{\rho}\vartheta
_{\rho\lambda}$ due the completely antisymmetric nature of $T.$

\subsection{Antisymmetric torsion and fermionic structure of the spacetime}

We know that [8]%

\begin{equation}
\Gamma_{\ \mu\lambda}^{\rho}=\left\{  _{\ \mu\lambda}^{\rho}\right\}
+g^{\rho\nu}\left(  T_{\mu\lambda\nu}+T_{\lambda\nu\mu}+T_{\nu\mu\lambda
}\right)  \tag{107}%
\end{equation}
where $\Gamma_{\ \mu\lambda}^{\rho}$ are he coefficients of the G-connection
and $\left\{  _{\ \mu\lambda}^{\rho}\right\}  $ denotes the coefficients of
the Levi-Civita connection whose covariant derivative is denoted by
$\overset{\circ}{\nabla}_{\lambda}.$From (105) we make the link between the
fermionic structure of the fundamental geometry of the manifold and the
torsion tensor%
\begin{align}
\nabla_{\left[  \nu\right.  }\sigma_{\left.  \lambda\mu\right]  }  &
=0\Rightarrow\tag{108}\\
\frac{1}{2}\partial_{\left[  \nu\right.  }\sigma_{\left.  \lambda\mu\right]
}  &  =T_{\left[  \nu\mu\right.  }^{\rho}\sigma_{\rho\left.  \lambda\right]  }
\tag{109}%
\end{align}
A particular simplest solution for T arises when the torsion tensor is totally
antisymmetric [9]%
\begin{equation}
T_{\mu\lambda\nu}=T_{\left[  \mu\lambda\nu\right]  } \tag{110}%
\end{equation}
in order that the equivalence principle be obeyed[5,9,10]. In this case, as we
shown already in [1,2,9], we have%
\begin{equation}
T_{\mu\lambda\nu}=\varepsilon_{\mu\lambda\nu\rho}h^{\rho} \tag{111}%
\end{equation}
where the axial vector $h^{\rho}$ is still to be determined. As will be clear
soon, is useful to put for d dimensions[9]%
\begin{equation}
h^{\rho}=\frac{1}{\sqrt{w}}J_{\lambda}^{\rho}P^{\lambda} \tag{112}%
\end{equation}
where $P^{\lambda}$ \ is the\textit{ generalized momentum vector. }If
$d=4$,$w=6.$

Expression (109) can be simplified taking account on the symmetries of
$T_{\mu\lambda\nu}$ and the contraction with the fundamental tensor $J_{\tau
}^{\lambda}$%
\begin{equation}
T_{\lambda\mu\nu}=\frac{1}{w}J_{\lambda}^{\rho}\partial_{\left[  \nu\right.
}\sigma_{\left.  \rho\mu\right]  } \tag{113}%
\end{equation}

\subsection{About the equivalence principle (EP) and the antisymmetry of the
torsion tensor: a theorem}

As is well known, in order that experimental evidence forms the foundation of
the theory, the PE\ has to be imposed as well the foregoing symmetry principles.

Because the G-connection contains a torsion tensor by specific requirements,
is currently suspected that due this fact, the EP can be violated. Then a good
question naturally arises: what is the implication of PE as defined (or better
described in this context) by the G-geometry? Let us analyze specifically the question

i) the PE $\Rightarrow$ the tangent space $M_{p}$ is to be a Minkowski space,
then at $M_{p}$ we have%
\begin{equation}
\left(  g_{\mu\nu}\right)  _{p}=\eta_{\mu\nu}\text{ and \ \ \ \ }\left(
\partial_{\rho}g_{\mu\nu}\right)  _{p}=0 \tag{A1}%
\end{equation}
where $\eta_{\mu\nu}$ is the Minkowski metric.

ii) The coefficients of the affine general connection is given by (17) [8
p.141]%
\begin{equation}
\Gamma_{\ \mu\lambda}^{\rho}=\left\{  _{\ \mu\lambda}^{\rho}\right\}
+\overset{\equiv S_{\ \mu\lambda}^{\rho}}{\overbrace{g^{\rho\nu}\left(
T_{\mu\lambda\nu}+T_{\lambda\nu\mu}+T_{\nu\mu\lambda}\right)  }} \tag{A2}%
\end{equation}
where $T_{\nu\mu\lambda}$ is the torsion tensor and $S_{\ \mu\lambda}^{\rho}$
is the contortion.

iii) from%
\begin{align}
\nabla g  &  =0\text{ we have, however}\nonumber\\
\nabla_{\lambda}g_{\alpha\beta}  &  =\overset{\circ}{\nabla}_{\lambda
}g_{\alpha\beta}-T_{\ \lambda\alpha}^{\rho}g_{\rho\beta}-T_{\ \lambda\beta
}^{\rho}g_{\alpha\rho}=0 \tag{A3}%
\end{align}
which is valid at $p$ also.

iv) from (A1) and (A3) we obtain%
\begin{equation}
\left[  T_{\beta\lambda\alpha}+T_{\alpha\lambda\beta}\right]  _{p}=0 \tag{A4}%
\end{equation}
since (A1) said%
\begin{equation}
\left[  \overset{\circ}{\nabla}_{\lambda}g_{\alpha\beta}\right]  _{p}=0
\tag{A5}%
\end{equation}
v) The above relations have \textit{tensorial character }, for instance they
are valid in all coordinate systems (and in all points $p$), then
\begin{equation}
T_{\beta\lambda\alpha}=-T_{\alpha\lambda\beta} \tag{A6}%
\end{equation}
and%
\begin{equation}
\overset{\circ}{\nabla}_{\lambda}g_{\alpha\beta}=0 \tag{A7}%
\end{equation}
These equations show geometrically that the imposition of the PE\ implies the
following equivalence%
\begin{equation}
\left[  \nabla_{\lambda}g_{\alpha\beta}=0\text{ and }PE\right]
\Longleftrightarrow\left(  eqs.\text{A6 and A7}\right)  \tag{A8}%
\end{equation}
vi) But, from (A6) and (A2) we have that the torsion tensor has the full
antisymmetric property%
\begin{equation}
T_{\alpha\lambda\beta}=T_{\left[  \alpha\lambda\beta\right]  } \tag{A9}%
\end{equation}
With this Proof we conclude that: \textit{the full antisymmetry for the
torsion tensor is the result of imposition of the Equivalence Principle (EP)
on the spacetime structure. Is not as the result of a priori assumptions
concerning the hypotetic or possible physical meaning of the torsion tensor}.

\subsection{The G-invariance of the action}

As is well known, the Palatini principle has a twice role that is the
determining of the connection required for the spacetime symmetry as the field
equations. By means this principle, we were able to construct the action
integral $S$. This action $S$ necessarily need to yield the G-invariant
conditions (104-106) without prior assumption; and, the Einstein, Dirac and
Maxwell equations need to arise from $S$ as a causally connected closed
system. This equations will be generalized inevitably, so that causal
connections between them can be established. Our action fulfill the above
requirements, having account that the role of $f_{\mu\nu}$that enters
symmetrically with $g_{\mu\nu}$ in S, is linked with the fundamental tensor
$\vartheta_{\mu\nu}$of the previous Section denoting the dual of
$\vartheta_{\mu\nu}$ by%
\[
f_{\mu\nu}\equiv\frac{1}{2}\varepsilon_{\mu\nu\rho\sigma}\vartheta^{\rho
\sigma}=\ast\vartheta_{\mu\nu}%
\]
(where $\vartheta^{\mu\nu}$ is the inverse tensor to $\vartheta_{\mu\nu})$

The usual Euler-Lagrange equations from the action with the explicit
computation of the determinant in (d=4) of expression (8) that will help us in
order to compare the unitarian model introduced here(in the sense of
Eddington[see 1,2]) with the dualistic non abelian Born-Infeld model of [3],
takes the familiar form [3,1,2]%
\begin{equation}
S=\frac{b^{2}}{4\pi}\int\sqrt{-g}dx^{4}\left\{  \overset{\equiv\mathbb{R}%
}{\overbrace{\sqrt{\gamma^{4}-\frac{\gamma^{2}}{2}\overline{G}^{2}%
-\frac{\gamma}{3}\overline{G}^{3}+\frac{1}{8}\left(  \overline{G}^{2}\right)
^{2}-\frac{1}{4}\overline{G}^{4}}}}\right\}  \tag{114}%
\end{equation}%
\begin{equation}
G_{\mu\nu}\equiv\left[  \lambda^{2}\left(  g_{\mu\nu}+f_{\ \mu}^{a}f_{a\nu
}\right)  +2\lambda R_{\left(  \mu\nu\right)  }+2\lambda f_{\ \mu}%
^{a}R_{\left[  a\nu\right]  }+R_{\ \mu}^{a}R_{a\nu}\right]  \tag{115}%
\end{equation}%
\begin{equation}
G_{\nu}^{\nu}\equiv\left[  \lambda^{2}\left(  d+f_{\mu\nu}f^{\mu\nu}\right)
+2\lambda\left(  R_{S}+R_{A}\right)  +\left(  R_{S}^{2}+R_{A}^{2}\right)
\right]  \tag{116}%
\end{equation}
with (the upper bar on the tensorial quantities indicates traceless condition)%
\begin{equation}%
\begin{array}
[c]{lllll}%
R_{S}\equiv g^{\mu\nu}R_{\left(  \mu\nu\right)  } & ; & R_{A}\equiv f^{\mu\nu
}R_{\left[  \mu\nu\right]  } & ; & \\
&  &  &  & \\
\gamma\equiv\frac{G_{\nu}^{\nu}}{d} & ; & \overline{G}_{\mu\nu}\equiv
G_{\mu\nu}-\frac{g_{\mu\nu}}{4}G_{\nu}^{\nu} & ; & \overline{G}_{\rho}^{\nu
}\overline{G}_{\nu}^{\rho}\equiv\overline{G}^{2}\\
&  &  &  & \\
\overline{G}_{\lambda}^{\nu}\overline{G}_{\rho}^{\lambda}\overline{G}_{\nu
}^{\rho}\equiv\overline{G}^{3} &  & \left(  \overline{G}_{\rho}^{\nu}%
\overline{G}_{\nu}^{\rho}\right)  ^{2}\equiv\left(  \overline{G}^{2}\right)
^{2} &  & \overline{G}_{\mu}^{\nu}\overline{G}_{\lambda}^{\mu}\overline
{G}_{\rho}^{\lambda}\overline{G}_{\nu}^{\rho}\equiv\overline{G}^{4}\\
&  &  &  &
\end{array}
\tag{117}%
\end{equation}
where the variation was made with respect to the electromagnetic potential
$a_{\tau}$ as follows
\begin{equation}
\frac{\delta\sqrt{G}}{\delta a_{\tau}}=\nabla_{\rho}\left(  \frac
{\partial\sqrt{G}}{\partial f_{\rho\tau}}\right)  \equiv\nabla_{\rho
}\mathbb{F}^{\rho\tau}=0 \tag{118}%
\end{equation}
Explicitly
\begin{equation}
\nabla_{\rho}\left[  \frac{\lambda^{2}N^{\mu\nu}\left(  \delta_{\mu}^{\sigma
}\ f_{\ \nu}^{\rho}+\delta_{\nu}^{\sigma}\ f_{\ \mu}^{\rho}\right)
}{2\mathbb{R}}\right]  =0 \tag{119}%
\end{equation}
where $N^{\mu\nu}$ is given by
\begin{equation}
N^{\mu\nu}=g\left[  -\gamma^{2}G^{\mu\nu}-\gamma\left(  G^{2}\right)  ^{\mu
\nu}+\frac{\left(  G^{2}\right)  _{\mu}^{\mu}G^{\mu\nu}}{2}-\left(
G^{3}\right)  ^{\mu\nu}+\frac{4\gamma^{3}g^{\mu\nu}}{d}-\frac{\gamma\left(
G^{2}\right)  _{\mu}^{\mu}g^{\mu\nu}}{d}-\frac{\left(  G^{3}\right)  _{\mu
}^{\mu}g^{\mu\nu}}{3d}\right]  \tag{120}%
\end{equation}
The set of equations to solve for the action (13) in this particular case is
\begin{gather}
R_{\left(  \mu\nu\right)  }=\overset{\circ}{R}_{\mu\nu}-T_{\mu\rho
}^{\ \ \ \alpha}T_{\alpha\nu}^{\ \ \ \rho}=-\lambda g_{\mu\nu}\tag{19a}\\
R_{\left[  \mu\nu\right]  }=\nabla_{\alpha}T_{\mu\nu}^{\alpha}=-\lambda
f_{\mu\nu}\tag{19b}\\
\nabla_{\rho}\left[  \frac{\lambda^{2}N^{\mu\nu}\left(  \delta_{\mu}^{\sigma
}\ f_{\ \nu}^{\rho}+\delta_{\nu}^{\sigma}\ f_{\ \mu}^{\rho}\right)
}{2\mathbb{R}}\right]  =0 \tag{19c}%
\end{gather}
from this set, the link between T and f will be determined (f is not a priori
potential for the torsion T)

The key point now is eq. (112)%
\begin{align}
\overset{\circ}{R}_{\mu\nu}  &  =-\lambda g_{\mu\nu}+T_{\mu\rho}%
^{\ \ \ \alpha}T_{\alpha\nu}^{\ \ \ \rho}\tag{121}\\
&  =-\lambda g_{\mu\nu}+wh_{\mu}^{\ \ \ }h_{\nu}^{\ \ \ }=-\lambda g_{\mu\nu
}+P_{\mu}P_{\nu} \tag{122}%
\end{align}
then we can obtain, as in mass shell condition
\begin{equation}
P^{2}=m^{2}\Rightarrow m=\pm\sqrt{\overset{\circ}{R}+\lambda d} \tag{123}%
\end{equation}
Notice that there exists a link between the dimension of the spacetime and the
scalar "Einstenian" curvature $\overset{\circ}{R}$. Moreover, the curvature is
constrained to take definite values $\in$ $\mathbb{N}$ the natural number
characteristic of the dimension. By the other hand,, knowing that $\left\vert
\lambda\right\vert =d-1$ and accepting that the parameter $m$ $\in\mathbb{R}$
, the limiting condition on the physical values for the mass is $\overset
{\circ}{R}\geqslant(1-d)d$

Introducing the geometric product in above equation (e.g.:$\gamma^{\mu}%
\gamma^{\nu}P_{\mu}P_{\nu}=m^{2}$) plus the quantum condition: $P_{\mu
}\rightarrow\widehat{P}_{\mu}-e\widehat{A}_{\mu}$. we have%
\begin{equation}
\left[  \gamma^{\mu}\gamma^{\nu}\left(  \widehat{P}_{\mu}-e\widehat{A}_{\mu
}\right)  \left(  \widehat{P}_{\nu}-e\widehat{A}_{\nu}\right)  -m^{2}\right]
\Psi=0 \tag{124}%
\end{equation}
where $\Psi=\mathbf{u}+i\mathbf{v}$ given in (91,92). That is%
\begin{equation}
\left[  \gamma^{\mu}\left(  \widehat{P}_{\mu}-e\widehat{A}_{\mu}\right)
+m\right]  \left[  \gamma^{\nu}\left(  \widehat{P}_{\nu}-e\widehat{A}_{\nu
}\right)  -m\right]  u^{\lambda}=0 \tag{125}%
\end{equation}
which lead the Dirac equation
\begin{equation}
\left[  \gamma^{\mu}\left(  \widehat{P}_{\mu}-e\widehat{A}_{\mu}\right)
+m\right]  u^{\lambda}=0 \tag{126}%
\end{equation}
with m given by (123). Notice that this condition, in the Dirac case, is not
only to pass from classical variables to quantum operators, but in the case
that the action does not contains explicitly $\widehat{A}_{\mu}$, $h_{\mu}%
$remains without specification due the gauge freedom in the momentum. Applying
the geometric product to (124) is not difficult to see that%
\begin{gather}
\left[  \left(  \widehat{P}_{\mu}-e\widehat{A}_{\mu}\right)  ^{2}-m^{2}%
-\frac{1}{2}e\sigma^{\mu\nu}F_{\mu\nu}\right]  u^{\lambda}+\frac{1}{2}%
\sigma^{\mu\nu}R_{\rho\left[  \mu\nu\right]  }^{\lambda}u^{\rho}-\nonumber\\
-\frac{1}{2}e\sigma^{\mu\nu}\left(  \widehat{A}_{\mu}\widehat{P}_{\nu
}-\widehat{A}_{\nu}\widehat{P}_{\mu}\right)  u^{\lambda}=0 \tag{127}%
\end{gather}

It is interesting to see that

i) the above formula is absolutely general for the type of geometrical
Lagrangians involved containing the generalized Ricci tensor inside,

ii) for instance, the variation of the action will carry the symmetric
contraction of components of the torsion tensor (i.e. eq.(121)), then the
arising of terms as $h_{\mu}^{\ \ \ }h_{\nu}^{\ \ \ },$

iii) the only thing that changes is the mass (123) and the explicit form of
the tensors involved as $R_{\rho\left[  \mu\nu\right]  }^{\lambda},$
$F_{\mu\nu}$ etc., without variation of the Dirac general structure of the
equation under consideration,

iv) eq. (127) differs from that obtained by Landau and Lifshitz by the
appeareance of the last two terms: the term involving the curvature tensor is
due the spin interaction with the gravitational field (due torsion term in
$R_{\rho\left[  \mu\nu\right]  }^{\lambda})$ and the last term is the spin
interaction with the the electromagnetic and mechanical momenta,

v) expression (127) is valid for another vector $v^{\lambda}$, then is valid
for a bispinor of the form $\Psi=\mathbf{u}+i\mathbf{v,}$

vi) the meaning for a quantum measurement of the spacetime curvature is mainly
due by the term in (127) involving explicitly the curvature tensor.

The important point here is that the spin-gravity interaction term is so
easily derived as the spinors are represented as spacetime vectors whose
covariant derivatives are defined in terms of the G-(affine) connection. In
their original form the Dirac equations would have, in curved spacetime, their
momentum operators replaced by covariant derivatives in terms of
\textquotedblleft spin-connection\textquotedblright whose relation is not
immediately apparent.

\section{Dirac structure, electromagnetic field and anomalous gyromagnetic
factor}

The interesting point now is based in the observation that if we introduce
expression (19b) in (127) then%

\begin{gather}
\left[  \left(  \widehat{P}_{\mu}-e\widehat{A}_{\mu}\right)  ^{2}-m^{2}%
-\frac{1}{2}e\sigma^{\mu\nu}F_{\mu\nu}\right]  u^{\lambda}-\frac{\lambda}%
{d}\frac{1}{2}\sigma^{\mu\nu}f_{\left[  \mu\nu\right]  }u^{\lambda
}-\nonumber\\
-\frac{1}{2}e\sigma^{\mu\nu}\left(  \widehat{A}_{\mu}\widehat{P}_{\nu
}-\widehat{A}_{\nu}\widehat{P}_{\mu}\right)  u^{\lambda}=0 \tag{128}%
\end{gather}%
\begin{equation}
\left[  \left(  \widehat{P}_{\mu}-e\widehat{A}_{\mu}\right)  ^{2}-m^{2}%
-\frac{1}{2}\sigma^{\mu\nu}\left(  eF_{\mu\nu}+\frac{\lambda}{d}f_{\mu\nu
}\right)  \right]  u^{\lambda}-\frac{e}{2}\sigma^{\mu\nu}\left(  \widehat
{A}_{\mu}\widehat{P}_{\nu}-\widehat{A}_{\nu}\widehat{P}_{\mu}\right)
u^{\lambda}=0 \tag{129}%
\end{equation}
we can see clearly that if $\widehat{A}_{\mu}=ja_{\mu}$ (with $j$ arbitrary
constant), $F_{\mu\nu}=jf_{\mu\nu}$ the last expression takes the suggestive
form%
\begin{equation}
\left[  \left(  \widehat{P}_{\mu}-e\widehat{A}_{\mu}\right)  ^{2}-m^{2}%
-\frac{1}{2}\left(  ej+\frac{\lambda}{d}\right)  \sigma^{\mu\nu}f_{\mu\nu
}\right]  u^{\lambda}-\frac{e}{2}\sigma^{\mu\nu}\left(  \widehat{A}_{\mu
}\widehat{P}_{\nu}-\widehat{A}_{\nu}\widehat{P}_{\mu}\right)  u^{\lambda}=0
\tag{130}%
\end{equation}
with the result that the gyromagnetic factor have been modified to $2/\left(
j+\frac{\lambda}{ed}\right)  .$ Notice that in an Unified Theory with the
characteristics introduced here is reasonable the identification introduced in
the previous step $(F\leftrightarrows f)$in order that the fields arise from
the same geometrical structure.

The concrete implications about this important contribution of the torsion to
the gyromagnetic factor will be given elsewhere with great detail on the
dynamical property of the torsion field. Only we remark the following:

i) there exists an important contribution of the torsion to the gyromagnetic
factor that can have implications to the trouble of the anomalous momentum of
the fermionic particles,

ii) this contribution appear (taking the second equality of expression 19b),
as a modification on the vertex of interaction, almost from the effective
point of view;

iii) is quite evident that this contribution will justify probably the little
appearance of the torsion at great scale, because we can bounded the torsion
due the other well know contributions to the anomalous momenta of the
elementary particles (QED, weak, hadronic contribution, etc.),

iv) the form of the coupling spin-geometric structure coming from the first
principles, as the Dirac equation, not prescriptions,

v) then, from iii) how the covariant derivative works in presence of torsion
is totally determined by the G structure of the spacetime,

vi) the Dirac equation (128) (where was introduced the second part of the
equivalence (19b) coming from the equation of motion), said us that the vertex
was modified without a dynamical function of propagation. Then, other form to
see the problem treated in this paragraph is to introduce the propagator for
the torsion corresponding to the first part of the equivalence (19b). This
important possibility will be studied elsewhere [5].

\section{Space-time and structural cohomologies}

As is well know from the physical and mathematical point of view, the
cohomological interplay between the fields involved in any well possessed
geometrical and unified theory is crucial. This importance arises as a
consequence of the logical (and causal) structure of the physical fields
(sources, fields, conserved quantities) and not only as a mathematical play.
In the theory presented here, there exist two cohomological structures:
\textit{Spacetime cohomology} and \textit{structural cohomology}

The difference between them is that in the \textit{Spacetime cohomology }the
Dirac (fermionic) structure of the space time is not involved directly in the
relations between the fields involved. The main equations necessary for the
construction are\textit{ }
\begin{align}
\nabla_{\alpha}T_{\ \ \mu\nu}^{\alpha}  &  =-\lambda f_{\mu\nu}\tag{131}\\
d^{\ast}T  &  =-\lambda^{\ast}f=dh\text{ \ \ \ }\nonumber
\end{align}
being the interplay schematically as%
\begin{equation}%
\begin{tabular}
[c]{lllllll}
&  &  & $\!\!\ $\fbox{$T$} &  &  & \\
&  & $\overset{A_{-}}{\swarrow}\underset{A_{+}}{\nearrow}$ &  & $\underset
{B_{-}}{\searrow}\overset{B_{+}}{\nwarrow}$ &  & \\
& \fbox{$\ f$} & \ \ \ \ \ \ \ \ \  & $\underset{C_{-}}{\overset{\ \ \ C_{+}%
}{\leftrightarrows}}$ &  & \fbox{$h$} &
\end{tabular}
\ \ \ \ \ \ \ \ \ \ \tag{132}%
\end{equation}
where the operators are%
\begin{equation}%
\begin{array}
[c]{cc}%
A_{-}\equiv\left(  -1\right)  ^{d+1}\left(  -\lambda\right)  \ast%
{\textstyle\int}
\ast & A_{+}\equiv\left(  -\lambda\right)  ^{-1}\ast d\ast\\
B_{-}\equiv\left(  -1\right)  ^{d+1}\ast & B_{+}\equiv\ast\\
C_{-}\equiv-\lambda\int^{\ast} & C_{+}\equiv\left[  \left(  -1\right)
^{d+1}\left(  -\lambda\right)  \right]  ^{-1}\ast d\\
D_{-}\equiv\left(  -1\right)  ^{d+1}\ast d & D_{+}\equiv\left(  -1\right)
^{d+1}\ast%
{\displaystyle\int}
\\
E_{-}\equiv d & E_{+}\equiv%
{\displaystyle\int}
\\
G_{-}\equiv\left[  \left(  -1\right)  ^{d+1}\left(  -\lambda\right)  \right]
^{-1}\ast & G_{+}\equiv-\lambda\ast
\end{array}
\tag{133}%
\end{equation}
The Structural cohomology, in contrast, involve directly the fermionic
structure of the spacetime due that in the basic formulas $\vartheta_{\mu\nu}$
enters directly into the cohomological game, as is easily seen below%

\begin{equation}%
\begin{array}
[c]{ccccc}
&  & \fbox{$a$} &  & \\
& ^{E_{-}}\swarrow\nearrow_{E_{+}} &  & _{D_{+}}\nwarrow\searrow^{D_{-}} & \\
\fbox{$f$} &  & \underset{B_{+}}{\overset{B_{-}}{\overrightarrow
{\longleftarrow}}} &  & \fbox{$\vartheta$}\\
& ^{C_{+}}\nwarrow\searrow_{C_{-}} &  & _{A_{-}}\swarrow\nearrow^{A_{+}} & \\
&  & \fbox{$h$} &  & \\
&  & _{G_{-}}\uparrow\downarrow_{G_{+}} &  & \\
&  & \fbox{$a$} &  &
\end{array}
\tag{134}%
\end{equation}
Notice the important thing that, in this case clearly the degree of the
relations between the quantities involved are more fundamental that in the
previous case ( jerarquical sense).

\section{Concluding remarks}

In this chapter we make an exhaustive analysis of the model based in the
theory developed in early references of the authors. The simplest structure of
the spacetime described by this new theory make, beside the connection between
curvature and matter, the link between the torsion and the spin.

As was well explained through all this paper, the mechanism of rupture of
symmetry is the responsible that the geometrical Lagrangian can be written in
a suggestive Eddington-Born-Infeld like form. Three cases were treated from
the point of view of the solutions, depending on the form of torsion used:
totally antisymmetric (with torsion potential), not totally antisymmetric
("tratorial " type ), and with a torsion tensor with both characteristics. In
all the cases they were compared from the point of view of the obtained
solutions with the non dualistic model of reference [3], namely the
Non-Abelian Born-Infeld model.

In all these cases the (non-dualistic) unified model proposed here have deep
differences with the dualistic non-Abelian Born-Infeld model of our early
reference [3].

The first obvious difference come from a conceptual framework: the geometrical
action will provide, besides the spacetime structure, the matter-energy spin
distribution. This fact is the same basis of the unification: all the
(apparently disconnected) theories and interactions of the natural world
appears naturally as a consequence of the intrinsic spacetime geometry.

For the case of totally antisymmetric tensor torsion with torsion potential,
several points were answered and elucidated:

i) about the Hosoya and Ogura ansatz the natural question arising was:

why the identification of the isospin structure of the Yang-Mills field with
the space frame lead a similar physical situation that a non-dualistic unified
theory with torsion? The answer is: because at once such identification is
implemented, a potential torsion is introduced and the solution of the set of
equations is the consistency between the definition of the torsion tensor from
the potential and the Cartan structure equations [1,2].

ii) about the obtained solutions for the scale factor, the difference with our
previous work is precisely the particular form of the energy-momentum tensor
in the NABI case (in the UFT model presented here, there are not
energy-momentum tensor, of course): both solutions describe a
wormhole-instanton but the final form of the differential equations for the
scale factor are different, then the scale factor here has an exponentially
growing behavior, in sharp contrast to the wormhole solution from our previous
work with the \textquotedblleft dualistic\textquotedblright\ non-Abelian BI
theory. Also, for this particular value of the torsion, the wormhole tunneling
interpretation (in the sense of the Coleman' s mechanism) is fulfilled.

The contact point between the compared models, however, are the dynamical
equations that are very similar although the existence of a \textquotedblleft
current term\textquotedblright\ in the UFT model (cf. (45)) that not appears
in the NABI case. This fact was pointed out in an slightly different context
by N. Chernikov.

For the case of non-totally antisymmetric (tratorial type) the spacetime
structure was analyzed from the point of view of the interacting fields
arising from the same geometry of the space time and relaxing now the
condition of a totally antisymmetric torsion, then, the prior existence of an
antisymmetric 2-form potential for it. The precise results can be easily
enumerated as:

(i) from its $SL(2C)$ underlying structure: the notion of minimal coupling has
been elucidated and come naturally of the compatibility condition between the
gauge field structure of the antisymmetric part of the fundamental tensor and
the SL(2,C) structure of the base manifold,

(ii) trough exact cosmological solutions from this model, where the geometry
is Euclidean $R\otimes O\left(  3\right)  \sim R\otimes SU(2)$, the relation
between the space-time geometry and the structure of the gauge group was
explicitly shown,

(iii) this relation is directly connected with the relation of the spin and
torsion fields

From the point of view of the obtained solutions, a solution of this model was
explicitly compared with our previous ones and we find that:

(i) the torsion is not identified directly with the Yang Mills type strength field,

(ii) there exists a compatibility condition connected with the identification
of the gauge group with the geometric structure of the space-time: this fact
lead the identification between derivatives of the scale factor a with the
components of the torsion in order to allows the Hosoya-Ogura ansatz (namely,
the alignment of the isospin with the frame geometry of the space-time),

(iii) this compatibility condition precisely mark the fact that local gauge
covariance, coordinate independence and arbitrary space time geometries are
harmonious concepts and

(iv) of two possible structures of the torsion the \textquotedblleft
tratorial\textquotedblright\ form forbids wormhole configurations, leading
only, cosmological instanton space-time in eternal expansion.

For the general case , i.e. with torsion with totally antisymmetric and
tratorial parts, the full analysis was given in a clear manner in Section VI.
Here we point out that the Hosoya and Ogura anzats can be implemented as in
the previous cases, and, the most important, the fact that wormhole solutions
can be obtained for some particular cases. The solutions are asymptotically
flat, where appear vector and tensor integration constants that are
constrained in norm to bring physical consistency to the solution.

About the problem of the possibility of coexistence of the trace of the
torsion due the tratorial part and the axial vector from the totally
antisymmetric part of the torsion, we saw here that there are not problem in
the new theory: there are tratorial and antisymmetric torsion fields without contradictions.

The fact that in reference [4] the field equations of vacuum quadratic
Poincare gauge field theory (QPGFT) were solved for purely null tratorial
torsion, if well permit to express the contortion tensor for such a case as
(tratorial form, with notation of ref.[4])%
\[
K_{\lambda\mu\nu}=-2(g_{\lambda\mu}a_{\nu}-g_{\lambda\nu}a_{\mu})
\]
does not permit the coexistence with an axial torsion vector, as was clearly
shown by Singh in the beautiful paper [4]. The two points that lead such
discrepancy are:

i) the different theories described, not only in foundations but also because
one is unitarian and the other of [4] dualistic

ii) and the fact that the Newman-Penrose formulation was used in [4], that as
is well known such method works in a null tetrad.

\subsection{On the geometrical structure}

From the point of view of the concrete structure able to explain the content
of the bosonic and fermionic matter of the universe, the present paper is left
open-ended as many physical consequences need to be explored. Some words
concerning to the realization and the choice of the correct group structure of
the tangent space to M is that $G=L\left(  4\right)  \cap Sp\left(  4\right)
\cap K\left(  4\right)  $ preserves the boson and fermion symmetry
simultaneously without imply supersymmetry of the model . As we like to show
in a future work, the supergravitational extension of the model will be
discussed joint with the problem of it quantization, where the key point will
be precisely the group structure of the tangent space to the spacetime
manifold M. Here we conclude enumerating the main results concerning to the
basic structure of the Manifold supporting an Unified Field Theoretical model:

i) the simplest geometrical structure able to support the fermionic fields was
constructed based in a tangent space with a group structure $G=L\left(
4\right)  \cap Sp\left(  4\right)  \cap K\left(  4\right)  $

ii) then, the explicitly link of the fermionic structure with the torsion
field was realized and the Dirac type equation was obtained from the same
spacetime manifold

iii) notice that the matter was not included on the Geometrical Lagrangian of
the Unified theory presented here: only symmetry arguments (that will lead the
correct dynamical equations for the material fields arising from the same
manifold) need to allow the appearance of matter and this fact is not the
essence of the unification, of course (several references trying to include
matter into the Eddington "type" theories by hand without physical and
symmetry principles).

\subsection{On the energy concept}

1) On the equation%
\[
\overset{\circ}{R}_{\mu\nu}=-\lambda g_{\mu\nu}+T_{\mu\rho}^{\ \ \ \alpha
}T_{\alpha\nu}^{\ \ \ \rho}%
\]
notice that the concept here of the terms that arise as "energy-momentum" part
coming from the symmetric contraction of the torsion components is different
in essence to the concept coming from the inclusion of the energy-momentum
tensor in the Einstein theory. The conceptual framework that "matter and
energy curve the spacetime" implicitly carry the idea of some "embedding-like"
situation where the matter and energy are putted on some Minkowskian flexible
carpet and you see how it is curved under the "weight" of the "ball"
(matter+energy). Here, in the theory presented, the situation is that the
torsion terms (contributing as "energy momentum in above equation) arise from
the same geometry, then we have the picture as an unique entity: the interplay
fields-spacetime . the idea is the same as the solitonic vortex in the water.

This fact can be also interpreted as that the concept of force is introduced
due the torsion in the unified model, thing that is lost in the Einstein
theory [10] where the concept is that there are not force, but curvature only.

2) Some remarks on the general Hodge-de Rham decomposition of $h=h_{\alpha
}dx^{\alpha}.$

\begin{theorem}
if $h=h_{\alpha}dx^{\alpha}$ $\notin F^{\prime}\left(  M\right)  $ is a 1-form
on $M$, then there exist a zero-form $\Omega$, a 2-form $\alpha=A_{\left[
\mu\nu\right]  }dx^{\mu}\wedge dx^{\nu}$ and an harmonic 1-form $q=q_{\alpha
}dx^{\alpha}$ on $M$ that%
\[
h=d\Omega+\delta\alpha+q\rightarrow h_{\alpha}=\nabla_{\alpha}\Omega
+\varepsilon_{\alpha}^{\beta\gamma\delta}\nabla_{\beta}A_{\gamma\delta
}+q_{\alpha}%
\]

\end{theorem}

Notice that if even is not harmonic and assuming that $q_{\alpha}$ is a polar
vector, an axial vector can be added such that above expression takes the form%
\[
h_{\alpha}=\nabla_{\alpha}\Omega+\varepsilon_{\alpha}^{\beta\gamma\delta
}\nabla_{\beta}A_{\gamma\delta}+\varepsilon_{\alpha}^{\beta\gamma\delta
}M_{\beta\gamma\delta}+q_{\alpha}%
\]
where $M_{\beta\gamma\delta}$ is a completely antisymmetric tensor.

3) Notice the important fact that when the torsion is totally antisymmetric
tensor field, $-2\lambda f_{\mu\nu}$ takes the role of "current" for the
torsion field, as usually the terms proportional to the 1-form potential
vector $a_{\mu}$ acts as current of the electromagnetic field $f_{\mu\nu}$ in
the equation of motion for the electromagnetic field into the standard
theory:$\nabla_{\alpha}f_{\ \ \mu}^{\alpha}=J_{\mu}$ (constants absorbed into
the $J_{\mu}$). The interpretation and implications of this question will be
analyzed concretely in [5].

\section{Acknowledgements}

I am very grateful to Professor Yu Xin that introduce me into the subject of
the Unified Theories based only in geometrical concepts and the Mach
principle. Many thanks are given to Professors Yu. P. Stepanovsky, and A.
Dorokhov for their interest and discussions.This work is partially supported
by \ Brazilian Ministry of Science and Technology (MCT).

\section{References}

[1] D. J. Cirilo-Lombardo, Int J Theor Phys (2010) 49: 1288--1301. (and
references therein)

[2] D. J. Cirilo-Lombardo, Int J Theor Phys DOI 10.1007/s10773-011-0678-1 (and
references therein)

[3] D.J. Cirilo-Lombardo, Class. Quantum Grav. \textbf{22} (2005) 4987--5004
(and references therein)

[4] P Singh , Class. Quantum Grav.7 (1990) 2125.

[5] D. J. Cirilo-Lombardo, in preparation.

[6] Kobayashi, Shoshichi; Nomizu, Katsumi, \textit{Foundations of Differential
Geometry, Volume 1 and 2}, Wiley-Interscience,(1996).

[7] David Hestenes, \textit{Space-Time Algebra}, Gordon \& Breach,(1966).

[8] Schouten J., \textit{Ricci-Calculus, }Heidelberg,Springer,(1954).

[9] Yu Xin, \textquotedblleft General Relativity on Spinor-Tensor
Manifold\textquotedblright, in: \textquotedblleft Quantum Gravity - Int.
School on Cosmology \& Gravitation\textquotedblright, XIV Course. Eds. P.G. Bergman,

V. de. Sabbata \& H.J. Treder, pp. 382-411, World Scientific (1996).

[10] Yu Xin, private communication.

[11] K. Borchsenius, Gen. Relativ. Gravit. 7, 527 (1976); R. T. Hammond,
Class. Quantum Gravity \textbf{6}, L195 (1989); N. J. Poplawski, Int. J.
Theor. Phys.\textbf{ 49,} 1481 (2010).

[12]T. W. B. Kibble, J. Math. Phys. \textbf{2}, 212 (1961); D. W. Sciama, Rev.
Mod. Phys. \textbf{36}, 463 (1964); 36, 1103 (1964); F. W. Hehl, Gen. Relativ.
Gravit. 4, 333 (1973); 5, 491 (1974); F. W. Hehl, P. von der Heyde, G. D.
Kerlick, and J. M. Nester, Rev. Mod. Phys. \textbf{48}, 393 (1976).

\bigskip

\bigskip
\end{document}